\documentclass[11pt]{article}
\usepackage{latexsym}
\usepackage[pdftex]{graphicx}
\usepackage{color}
\usepackage{amsmath,amssymb}
\usepackage{cite}

\newcommand{\beq}{\begin{eqnarray}}
\newcommand{\eeq}{\end{eqnarray}}

\newcommand{\gsim}{\raisebox{-0.13cm}{~\shortstack{$>$ \\[-0.07cm]
     $\sim$}}~}
\newcommand{\lsim}{\raisebox{-0.13cm}{~\shortstack{$<$ \\[-0.07cm]
     $\sim$}}~}
\newcommand{\s}{\newline \vspace*{-3.5mm}}
\newcommand{\comment}[1]{}

\begin{document}

\renewcommand{\thefootnote}{\fnsymbol{footnote}}
\setcounter{page}{0}
\thispagestyle{empty}

\begin{titlepage}

\vskip-1.0cm

\begin{flushright}
\hfill CERN-PH-TH/2011-019 \\
\hfill ITP-UU-11/09 \\
\hfill KA-TP-07-2011 \\
\hfill SFB-CPP-11-12
\end{flushright}

\begin{center}

\vspace{1.7cm}

{\LARGE\bf 
Model-independent analysis of Higgs spin and CP properties in
the process \boldmath{$e^+ e^- \to t \bar t \Phi$}} 

\vspace{1.4cm}

{\bf R.M. Godbole\footnote{On leave of absence from: Center for High Energy Physics, Indian Institute of Science, Bangalore 560 012, India}$^{\,a,\,b}$}, {\bf C. Hangst$^{\,c}$}, {\bf
  M. M\"uhlleitner$^{\,c}$}, {\bf S.D. Rindani$^{\,d}$} and {\bf
  P. Sharma$^{\,d}$}\\ 

\vspace{1.2cm}

${}^{a}\!\!$
{\em {Theory Division, CERN TH-PH, CH-1211, Geneva 23, Switzerland}
}\\
${}^b\!\!$
{\em {Institute for Theoretical Physics and Spinoza Institute, Utrecht
  University, 3508 TD Utrecht, The Netherlands}
}\\
${}^c\!\!$
{\em {Institut f\"ur Theoretische Physik, Karlsruhe Institute of Technology, 76128 Karlsruhe, Germany}
}\\
${}^d\!\!$
{\em {Theoretical Physics Division, Physical Research Laboratory, Navrangpura, Ahmedabad 380 009, India}
}\\

\end{center}

\vskip 2cm

\begin{abstract}
\noindent
In this paper we investigate methods to study the $t\bar{t}$ Higgs
coupling. The spin and CP properties of a Higgs boson are analysed in a
model-independent way in its associated production with a $t\bar{t}$ pair
in high-energy $e^+e^-$ collisions. We study the prospects of
establishing the CP quantum numbers of the Higgs boson in 
the CP-conserving case as well as those of determining the CP-mixing
if CP is violated. We explore in this analysis the combined use of
the total cross section and its energy dependence, the polarisation
asymmetry of the top quark and the up-down asymmetry of the antitop
with respect to the top-electron plane. We find that combining
all three observables remarkably reduces the error on the
determination of the CP properties of the Higgs Yukawa coupling. Furthermore,
the top polarisation asymmetry and the ratio of cross sections at different
collider energies are shown to be sensitive to the spin of the
particle produced in association with the top quark pair.
\end{abstract}

\end{titlepage}

\renewcommand{\thefootnote}{\arabic{footnote}}
\setcounter{footnote}{0}
%
\section{Introduction \label{sec:Intro}}
One of the major goals of the LHC is to probe if the Higgs mechanism
\cite{Higgs,Goldstone,Gunion} is at the origin of electroweak symmetry
breaking (EWSB). Whereas at the LHC \cite{lhc} a Standard Model (SM)
Higgs boson can be found in the whole canonical mass range and first
information on its properties will be available, the clean environment of the
International Linear Collider (ILC) \cite{ilc} would be needed in
order to determine the particle properties with sufficient precision. The
interplay of the information available from both colliders  will provide us with a 
clearer picture of the dynamics behind the creation of the masses of
the fundamental particles \cite{lhcilc}. This requires the
determination of the spin $J$ and the CP quantum numbers of the observed 
state, the measurement of its couplings to gauge bosons and fermions
and finally the extraction of the trilinear and quartic Higgs
self-couplings \cite{Djouadi:1999rca} to reconstruct the Higgs
potential itself. \s

The SM Higgs boson is predicted to be CP-even and have spin zero, 
{\it i.e.} $J^{CP}=0^{++}$. Should extensions beyond the SM (BSM) be realised, the
existence of more than one spin zero particle is possible. Some of
these particles can have CP-odd properties or even be states with no
definite CP quantum number. Supersymmetric theories
\cite{susy,susy-book} for example require the introduction of at least
two complex Higgs doublets. In the Minimal Supersymmetric Extension of
the Standard Model (MSSM) \cite{Gunion,Djouadi:2005gj} this leads
after electroweak symmetry breaking to five physical Higgs states,
out of which three are neutral, two CP-even and one  CP-odd. Should CP
violation beyond the one of the SM be provided by the Higgs sector,
this would imply Higgs states with mixed CP properties. CP
violation is one of the conditions to explain the Baryon Asymmetry in
the Universe \cite{sakharov}. A study of the CP properties of the Higgs boson
can hence already give some clues to BSM physics~\cite{Accomando:2006ga}. 
Such a study at the current and future colliders\cite{Godbole:2004xe}
would have to establish the CP properties of the observed spin 0
particle(s) and determine the amount of CP-mixing in case of CP
violation. This can be done by establishing the tensor
structure of the Higgs couplings to $Z$ or $W$ pairs 
and of the $f \bar f \Phi$ couplings, where $\Phi$ denotes a
general $J=0$ state. Any deviation from the SM prediction can then be
interpreted in the framework of a given model. 
\s

In the SM, the largest Higgs boson couplings are the ones to the
heaviest fermions and to a massive $W$ or $Z$ gauge boson
pair. Therefore Higgs production via these couplings or its decays
into a heavy $f \bar f$ or  $WW,ZZ$ pair provide the best processes to
obtain information on its spin and parity. The couplings to a pair of
photons or gluons, which proceed through loops of these particles, can
be used, too. A plethora of observables, constructed out of the kinematic 
variables of the different particles involved in the production and 
decay of the Higgs boson, can provide this information. The strategy
for a model-independent test and analysis is to assume the most 
general form of the relevant Higgs couplings consistent with Lorentz
invariance and gauge invariance and explore how these couplings can be
constrained by using the mentioned observables. This is usually
possible by exploiting properties of kinematical distributions such as
threshold effects and angular correlations in Higgs decays into $W$ or
$Z$ pairs or production in $\gamma \gamma$ fusion, without 
\cite{Grzadkowski:1992sa,Barger:1993wt,Kramer:1993jn,skjold,higgsdecay,Accomando:2006ga}
and with the inclusion of CP-violating effects
\cite{Chang:1993jy,Skjold:1994qn,Asakawa:2000jy,Godbole:2002qu,Choi:2004kq,Niezurawski:2004ga,Godbole:2006eb,Godbole:2007cn,DeRujula:2010ys}. In
the decay channel into a $Z$ boson pair even small signal samples, as
available at the moment of Higgs discovery at the LHC, might be
sufficient to draw first conclusions on the CP properties of the Higgs
boson \cite{DeRujula:2010ys}. The Higgs production processes through
Higgs-radiation and gauge boson 
fusion in $e^+e^-$ collisions also provide observables to extract
spin parity informations of the Higgs boson as well as the  couplings
of the Higgs boson with a  gauge boson pair 
\cite{Godbole:1982mx,Kramer:1993jn,Hagiwara:1993sw,Miller:2001bi},
with CP violation included in 
\cite{Chang:1993jy,Gounaris:1995mx,Biswal:2008tg,Rao:2007ce}. In addition, 
the azimuthal angular distribution of the two outgoing forward tagging
jets in weak boson fusion and in gluon fusion initiated processes can
be used \cite{Plehn:2001nj,gfusioninit,multijet}. In vector
boson fusion the CP nature of the Higgs $VV$ ($V= W,Z$) 
coupling is probed, whereas gluon fusion provides this information for
the Higgs coupling to a top quark pair. It is found that a CP-even
Higgs boson can be clearly distinguished from a CP-odd Higgs state.\s

The coupling of a pseudoscalar Higgs boson to a $VV$ pair is always 
loop induced and hence suppressed with respect to the tree level coupling of a
scalar state. As a result, the best among the different methods suggested to 
study the CP-mixing are those which use the couplings of a $J=0$ state
to a pair of photons
\cite{Grzadkowski:1992sa,Asakawa:2000jy,Godbole:2002qu,Choi:2004kq,Niezurawski:2004ga,Godbole:2006eb}
at the  photon collider option \cite{Badelek:2001xb} of the ILC, and to the 
heavy fermions $t$ or $\tau$
\cite{Grzadkowski:1995rx,BarShalom:1995jb,Bernreuther:1993df,Bernreuther:1998qv,Gunion:1996vv,Gunioneetth,Huang:2001ns,Khater:2003wq,Bower:2002zx,BhupalDev:2007is}
at the LHC and the ILC. The advantage is that these particles couple
democratically to the CP-even and the CP-odd components of the Higgs
boson. As the photon collider option is the most remote one to be
realized, the study of the associated production of a $J=0$ state with
a $f \bar f$ pair at the LHC or ILC or the analysis of the decays of a heavy Higgs
produced via gluon fusion into a $f \bar f$ pair are the most
promising candidates to get unambiguous information on the CP
properties of a state with indefinite CP quantum numbers. \s 

The top quark is the heaviest fermion discovered so far and, providing
the largest Yukawa coupling, plays a special role. At a future ILC
Higgs boson production in association with a top quark pair,
$e^+e^-\to t\bar{t} \Phi$ 
\cite{Gaemers:1978jr,Dittmaier:1998dz,Dittmaier:2000tc,Zhu:2002iy,You:2003zq}
leads to sufficiently high rates \cite{Juste:1999af} and can hence be used to
extract CP information  
\cite{BarShalom:1995jb,Gunioneetth,BhupalDev:2007is,Huang:2001ns} by 
exploiting angular correlations and/or the polarisation of heavy
fermions \cite{BhupalDev:2007is,Huang:2001ns}. These can also be
exploited in the Higgs decays to heavy fermions
\cite{Grzadkowski:1995rx}. 
Reference \cite{Gunioneetth} employs an optimal observable technique
to show that the discrimination between a CP-even and a CP-odd case at a 
high level of statistical significance is possible with low luminosity, and that a 
non-zero CP-violating coupling can be determined at the $1\sigma$
level with $\int {\cal{L}} = 100$ fb$^{-1}$. 
In Ref.~\cite{BhupalDev:2007is} the extraction of the Higgs CP quantum
numbers  in $t\bar{t} \Phi$ production at a future ILC through the
measurement of the total cross section and its energy dependence as
well as of the top polarisation asymmetry has been discussed. In the current
paper we extend the analysis in two ways. Firstly, we allow for more
than two neutral spin $0$  states while writing the most general 
CP-violating $f \bar f \Phi$ vertex. Secondly, we investigate possible
sensitivities to probe this general $f \bar f \Phi$ vertex in the
combined study of three different observables. The observables we use
are the total cross section $\sigma$ and its energy dependence,
which has been shown to exhibit an unambiguously different behaviour
for CP-even and CP-odd Higgs bosons, the top quark polarisation
asymmetry $P_t$ and the up-down asymmetry $A_\phi$ of the antitop
quark with respect to the top-electron plane. The latter can directly
probe CP violation. We also discuss how well the energy dependence of
the cross section and the top polarisation asymmetry can differentiate
between the case of a spin 1 particle produced in association with a
$t \bar t$ pair and that of a spin 0 particle. \s 

The outline of the paper is as follows. In section 2 we will discuss
the observables and their individual sensitivities to the Higgs CP
properties. Then the combination of all observables will be investigated
with respect to an improvement in the determination of the CP quantum
numbers. In section 3 the associated production of a spin 1 particle with a top
quark pair will be discussed to investigate to what extent the total
cross section and the top polarisation asymmetry can help to
distinguish a spin 1 from a spin 0  state. We will conclude in section 4.

 \section{\label{sec:observables} The observables}
In this section we will discuss the observables for the production of an
arbitrary CP-violating Higgs boson $\Phi$ in association with a
top-quark pair at a future $e^+e^-$ linear collider,
\beq
e^+ e^- \to t\bar{t} \Phi \:.
\eeq
The value of the total cross section $\sigma$ and its energy dependence
exhibit sensitivity to the CP nature of the 
Higgs boson and allow for a distinction between a CP-odd and a
CP-even Higgs boson. The top quark polarisation asymmetry $P_t$ serves as
a further observable. Both $\sigma$ and $P_t$ are CP-even and hence do
not test CP violation. The $t\bar{t}\Phi$ production
process, however, exhibits already at tree-level CP violation in the
interference between the diagrams where $\Phi$ is emitted from the top
(antitop) and where it is emitted from the $Z$ boson
\cite{BarShalom:1995jb}. The up-down asymmetry $A_\phi$ projects out
this interference term and thus tests CP violation. 
In the following we will discuss the observables $\sigma$, $P_t$ and
$A_\phi$ in detail as well as their sensitivity to the Higgs
CP properties. We will examine if the longitudinal polarisation of the
initial $e^\pm$ beams can help to improve the sensitivity. For the
degree of polarisation, the standard ILC values of $P_{e^-}=-0.8$ and
$P_{e^+}=0.6$ will be assumed. Positive values of $P_{e^-}, P_{e^+}$
correspond to right-handed polarisation. Finally, we will combine all
observables to derive the errors on the CP parameters in a general
CP-violating $t\bar{t} \Phi$ coupling by performing a $\chi^2$ test. 

\subsection{The total cross section}
\begin{figure}[!h]
\begin{center}
\includegraphics[width=1.1\linewidth,bb=73 699 600 735]{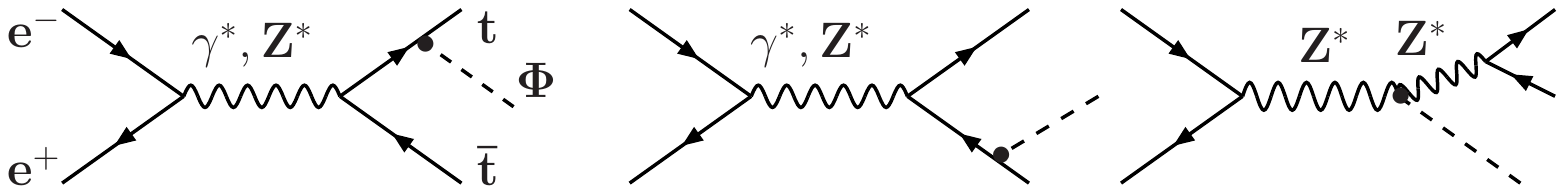} 
\vspace*{0.1cm}
\caption[]{\label{fig:feyndiag} Feynman diagrams contributing to the
  associated production of a Higgs boson $\Phi$ with a top quark pair.}
\end{center}
\end{figure}

The associated production of a SM Higgs boson with a top quark pair,
$e^+e^- \to t\bar{t} H$, can be measured with an accuracy of ${\cal
  O}(10)$\% for $M_H \lsim 200$ GeV
\cite{Juste:1999af}. At tree level
\cite{Gaemers:1978jr}, it proceeds through the Higgs radiation off 
the $t,\bar{t}$ lines and a diagram with the Higgs boson produced in
association with a $Z$ boson which then splits into a $t\bar{t}$ pair, {\it cf.}
Fig. \ref{fig:feyndiag}. This diagram only contributes with a few percent
as long as $\sqrt{s} \le 1$ TeV. In fact, the bulk of
the cross section stems from the splitting of the virtual photon into
$t\bar{t}$. \s

In models with more than one Higgs boson as {\it e.g.} a general
2-Higgs doublet model (2HDM) or supersymmetric extensions of the SM,
there are additional diagrams with a CP-odd (even) Higgs boson splitting into
$t\bar{t}$ for (pseudo-)scalar Higgs production. In CP-violating
models both Higgs bosons are CP-mixed states, one of them splitting
into a top quark pair. This contribution is in general small unless
Higgs decays into $t\bar{t}$ are kinematically allowed. We assume 
here that by the time the CP quantum numbers of Higgs bosons
will be tested, all Higgs bosons will have been discovered and their
masses will have been determined. By applying appropriate cuts on the
invariant mass of the $t\bar{t}$ pair we can hence
safely neglect these additional diagrams.\footnote{In case the Higgs
  bosons are not discovered because of suppressed couplings to gauge
  bosons or too heavy masses, the process is not affected by these
  additional Higgs bosons either. Their influence is negligible
  in the first case, and they won't play any role in the second
  case at sufficiently low c.m. energy.}  \s

In order to discuss a general CP-mixed Higgs state $\Phi$ we
parametrize the $t\bar{t} \Phi$ coupling by
\beq
C_{tt\Phi} = -i \frac{e}{\sin \theta_W} \frac{m_t}{2M_W}
(a+ib\gamma_5)  \equiv -i g_{ttH} (a+ib \gamma_5)\;,
\eeq
where $\theta_W$ denotes the Weinberg angle, $m_t$ the top quark mass
and $M_W$ the $W$ boson mass.
The coefficients $a$ and $b$ are assumed to be real and in the SM are
given by $a=1,b=0$. A pure pseudoscalar coupling is provided by
$a=0,b\ne 0$. A coupling to a Higgs state with indefinite CP quantum
numbers is realized if simultaneously $a \ne 0$ and $b \ne 0$. The
exact values of these parameters depend on the model under
consideration. In the MSSM for example the Higgs couplings to
up-(down-)type quarks are suppressed (enhanced) for large values of
$\tan\beta$, the ratio of the vacuum expectation values of the two
complex Higgs doublets \cite{Gunion,Boos:2002ze}. From the
experimental side the upper bounds on the electric dipole moments of
the neutron and the electron provide important constraints on
non-standard Higgs sector CP violation \cite{Accomando:2006ga}. Within
a general 2HDM with maximal CP violation, values with $|ab| \lsim 2$
are in accordance with low-energy constraints
\cite{Hayashi:1994ha,Bernreuther:1998qv}. We work in a
model-independent approach and in the following let $a$ and $b$ vary
between $-1$ and 1, if not stated otherwise. \s 

For the calculation of the total cross section the $ZZ\Phi$ coupling
will be needed, too. It is parametrized in terms of the SM coupling
$g_{ZZH}$ by a parameter $c$, {\it i.e.}
\beq
g^{\mu\nu}_{ZZ\Phi} = 
-ic \, \frac{e M_Z}{\sin\theta_W \cos\theta_W}
g^{\mu\nu} \equiv -ic \, g_{ZZH} \, g^{\mu\nu}\;.
\eeq
The $ZZ\Phi$ coupling and hence the parameter $c$ will be determined
from other channels \cite{lhc,ilc,zzphilhc,zzphiilc} so that
the analysis for the determination of the Higgs CP properties can be
performed for a fixed value of $c$. In order to reduce the number of
free parameters, we will choose $c=-a$ in our
analysis. This is a reasonable choice as $c=0$ in case of a CP-odd
Higgs boson and $|c| \le 1$ in general\footnote{This follows from a
  sum rule for the Higgs gauge coupling
  \cite{Gunion:1990kf,Mendez:1991gp} in models with a Higgs 
  sector containing only Higgs singlets and doublets.}. 
For $a=1$ the $ZZ\Phi$ coupling then takes the SM form. In principle,
$ZZ$ can also couple to the CP-odd part of the Higgs boson 
\cite{Mendez:1991gp,oddzz,oddzz1}.  
This coupling is, however, zero at tree-level and only generated
through tiny loop corrections. Therefore we consider only $ZZ\Phi$
couplings without contributions involving the CP-odd part of $\Phi$. \s 

Neglecting the small contribution of the diagram involving the
$Z Z \Phi$ vertex, the differential cross section with respect to the scaled
energies $x_{1,2}=2 E_{t, \bar t}/\sqrt{s}$, can be cast into the form
\begin{eqnarray}
\frac{d\sigma}{dx_1 dx_2} =
\frac{\alpha^2}{4 \pi s} \bigg\{ \hspace*{-0.3cm} 
&\bigg[ & \hspace*{-0.4cm} 
Q_e^2 Q_t^2 +\frac{( {v}_e^2  
+ {a}_e^2) ({v}_t^2 + {a}_t^2)}{(1- h_z)^2} 
+ \frac{2 Q_e Q_t {v}_e {v}_t}{1-h_z} \bigg] \left(a^2 F_1^H + b^2 F_1^A
  \right) \nonumber \\ 
\hspace*{-0.3cm} &+& \hspace*{-0.3cm} 
\frac{{v}_e^2+ {a}_e^2}{(1- h_z)^2} a_t^2 \left(a^2 F_2^H + b^2
  F_2^A \right) \bigg\} \, g_{ttH}^2 \;,
\label{ttHxsection}
\end{eqnarray}
where $\alpha^{-1}\!=\!\alpha^{-1}(s)\!\sim\!128$, $h_z\!=\!M_Z^2/s$ and
${v}_{f}\!=\!(2I^{3L}_{f}-4 Q_{f}s^2_W) /(4s_Wc_W) , {a}_{f}\!
=\!2I^{3L}_{f}/(4s_Wc_W)$ are the usual $Zf\bar{f}$ couplings given in
terms of the fermion charge $Q_{f}$ and the third component of the
weak isospin $I^{3L}_{f}$ ($f=t,e$). We have introduced the 
short-hand notation $s_W=\sin\theta_W, c_W = \cos\theta_W$. The
expressions for the form factors  $F^{H,A}_{1,2}$ for scalar and
pseudoscalar Higgs bosons are given in Refs.~\cite{Gaemers:1978jr}. Note
that in the spin averaged matrix element terms proportional to the
combination $ab$ of the CP-even and CP-odd coupling
parameters in the Yukawa coupling do not appear. \s

\begin{figure}[t]
\centering
\vspace*{-0.85cm}
\includegraphics[width=11cm]{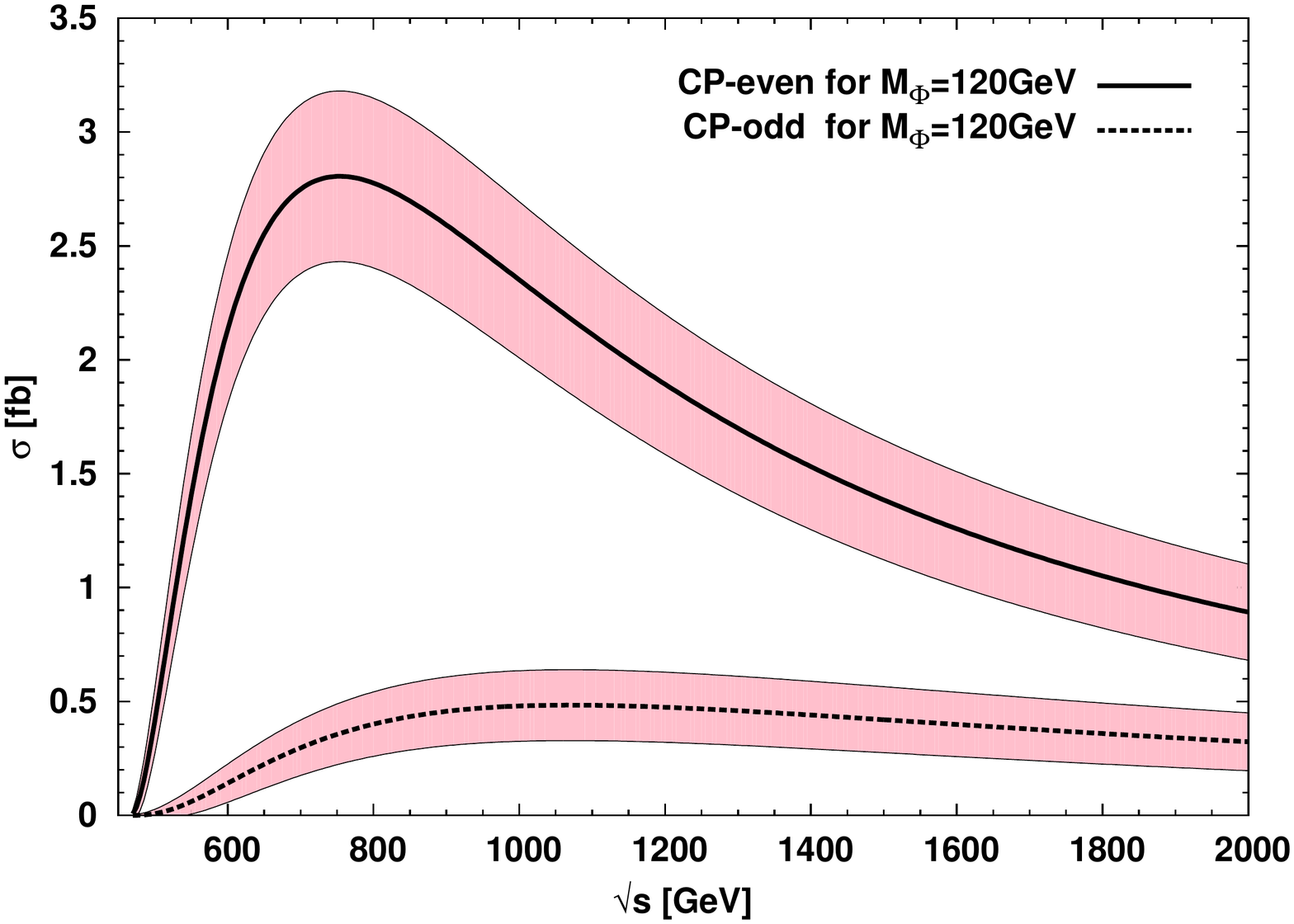}
\caption{Total cross section in fb for SM (full) and purely 
  CP-odd (dashed) Higgs production in association with a $t\bar{t}$
  pair as function of the c.m. energy for $M_\Phi=120$ GeV and
  unpolarised $e^\pm$ beams. The pink (grey) band indicates the
  statistical error at $\pm 5\sigma$ with $\int {\cal L} = 500$
  fb$^{-1}$.} 
\label{fig:rise500}
\end{figure}

The total cross section is a CP-even observable and not sensitive to
possible CP violation. Its threshold rise, however, is strikingly
different for the scalar and pseudoscalar case as has been shown in
Ref.~\cite{BhupalDev:2007is}. Parametrizing the deviation $\rho$
from the threshold by 
\beq
\rho = 1- \frac{2m_t}{\sqrt{s}} - \frac{M_\Phi}{\sqrt{s}} \;,
\eeq
the latter shows a softer dependence on $\rho$, which is given by $\sim
\rho^3$. The rise for a purely CP-even Higgs boson $H$ on the other hand
is $\sim \rho^2$. The threshold rise can hence be exploited to
distinguish a pseudoscalar from a scalar Higgs boson. Taking into
account only statistical fluctuations, for a 120 GeV Higgs boson an
integrated luminosity of 500 fb$^{-1}$ is sufficient to distinguish
the purely pseudoscalar from the SM case at 5$\sigma$ confidence level,
as can be inferred from Fig.~\ref{fig:rise500}. According to our coupling
parametrization, here and in the following SM refers to $a=1,b=0$ in
the $t\bar{t}\Phi$ coupling and purely pseudoscalar refers to
$a=0,b=1$, if not stated otherwise. \s 

In the following we discuss to what extent the value of the total
cross section can be exploited to extract the CP properties of the
Higgs boson. The cross section is subject to higher-order (SUSY-) QCD 
\cite{Dittmaier:1998dz,Dittmaier:2000tc,Zhu:2002iy} and electroweak (EW)
\cite{You:2003zq} corrections. The QCD corrections can be significant
for collider energies near the threshold. In the continuum, for
$\sqrt{s}=1$ TeV they are of moderate size for both the scalar and
pseudoscalar Higgs boson. The EW corrections can reach about $10\%$.
For a precise determination of the top Yukawa coupling the
corrections have to be taken into account. In this analysis we neglect
their influence in a first approximation. We will supplement our
analysis later with the use of the polarisation asymmetry of the top
quark which we expect to be less sensitive to such corrections. \s

\begin{figure}[t]
\centering
\includegraphics[width=0.45\textwidth]{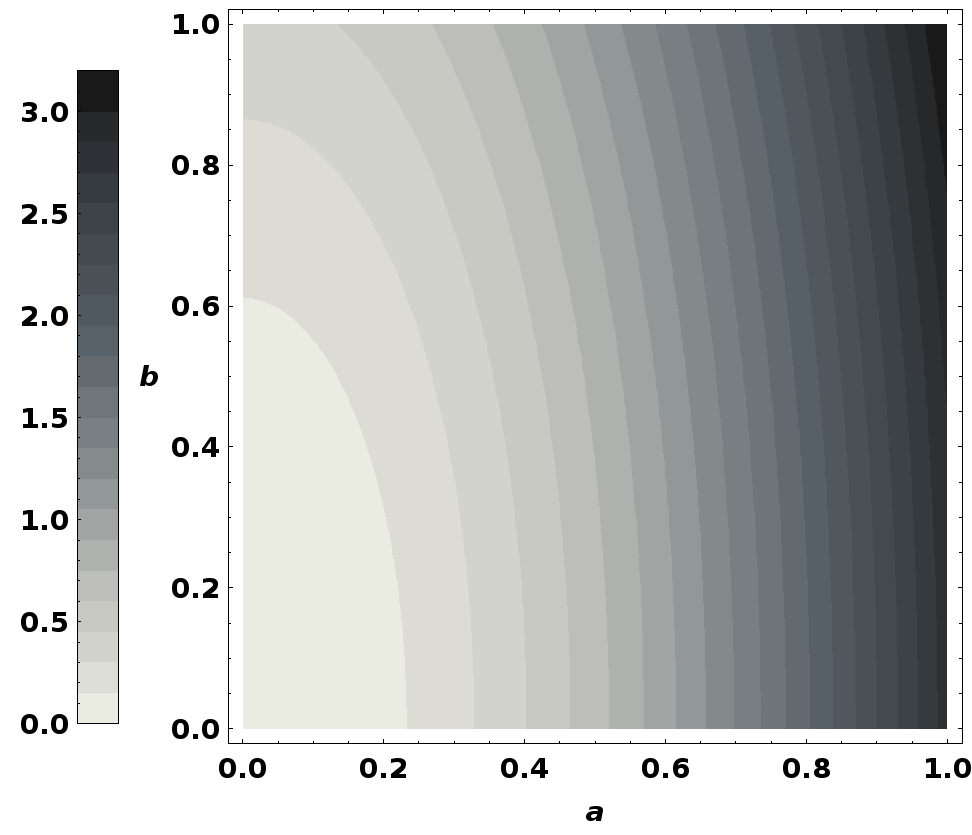}
\hspace*{0.6cm}
\includegraphics[width=0.45\textwidth]{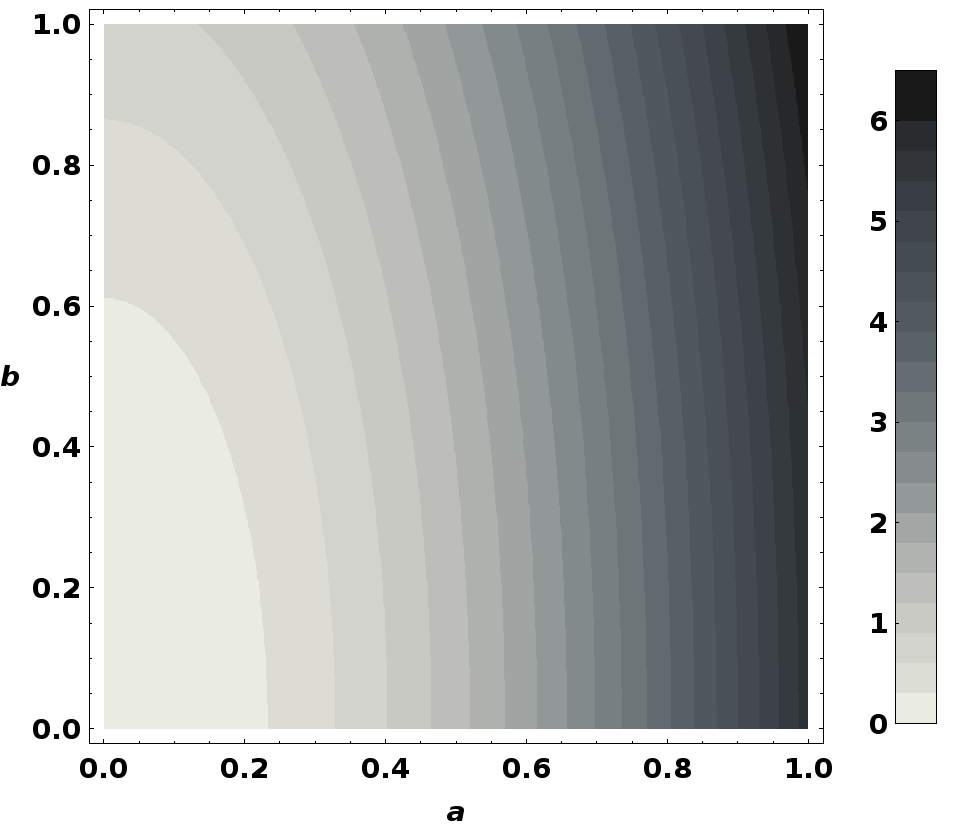}
\caption{\label{fig:sig0contour} Contour plot of the total cross section
  $\sigma(e^+e^-\to t\bar{t}\Phi)$ in fb in the $a-b$ parameter plane for
  $M_H=120$ GeV and $\sqrt{s}=800$ GeV with unpolarised (left) and
  polarised (right) $e^\pm$ beams. The grey code indicates the
  value of the cross section.}
\vspace*{0.2cm}
\end{figure}

The total $t\bar{t}\Phi$ cross section values are shown for
unpolarised $e^\pm$ beams in Fig. \ref{fig:sig0contour} (left) as contours
in the $a-b$ plane. We have chosen $M_\Phi=120$ GeV and the
c.m. energy $\sqrt{s}=800$ GeV. Note that here and in the following
the diagram with the $ZZ\Phi$ coupling is always taken into account in
the numerical analyses.
Since the cross section depends quadratically on $a,b$ (for our choice
of $c=-a$) the contour lines are symmetric with respect to the sign of
$a$ and $b$ so that only results for positive values of $a$ and $b$
are shown. The cross section decreases with decreasing values of
$a,b$. 
The SM value of the cross section amounts to 2.78 fb. The cross
section for a purely CP-odd Higgs boson is with 0.4 fb much smaller
than the one for a SM Higgs boson, see also Fig.~\ref{fig:rise500}.
Furthermore,
the dependence on $b$ is rather flat in contrast to the dependence on
$a$. We hence expect the cross section to be mostly sensitive to
$a$. Polarisation of the initial $e^\pm$ beams can help to increase
the cross section for our choice of $P_{e^\pm}$, as shown in
Fig.~\ref{fig:sig0contour} (right). It increases the cross section by
about a factor of 2, independent of the CP nature of the Higgs
boson. In the Appendix the replacements are listed, which have to be
applied in the formula of the cross section in case of initial beam polarisation. 
\s

\begin{figure}[t]
\vspace*{-0.5cm}
\centering
\hspace*{-0.3cm}
\includegraphics[width=16cm]{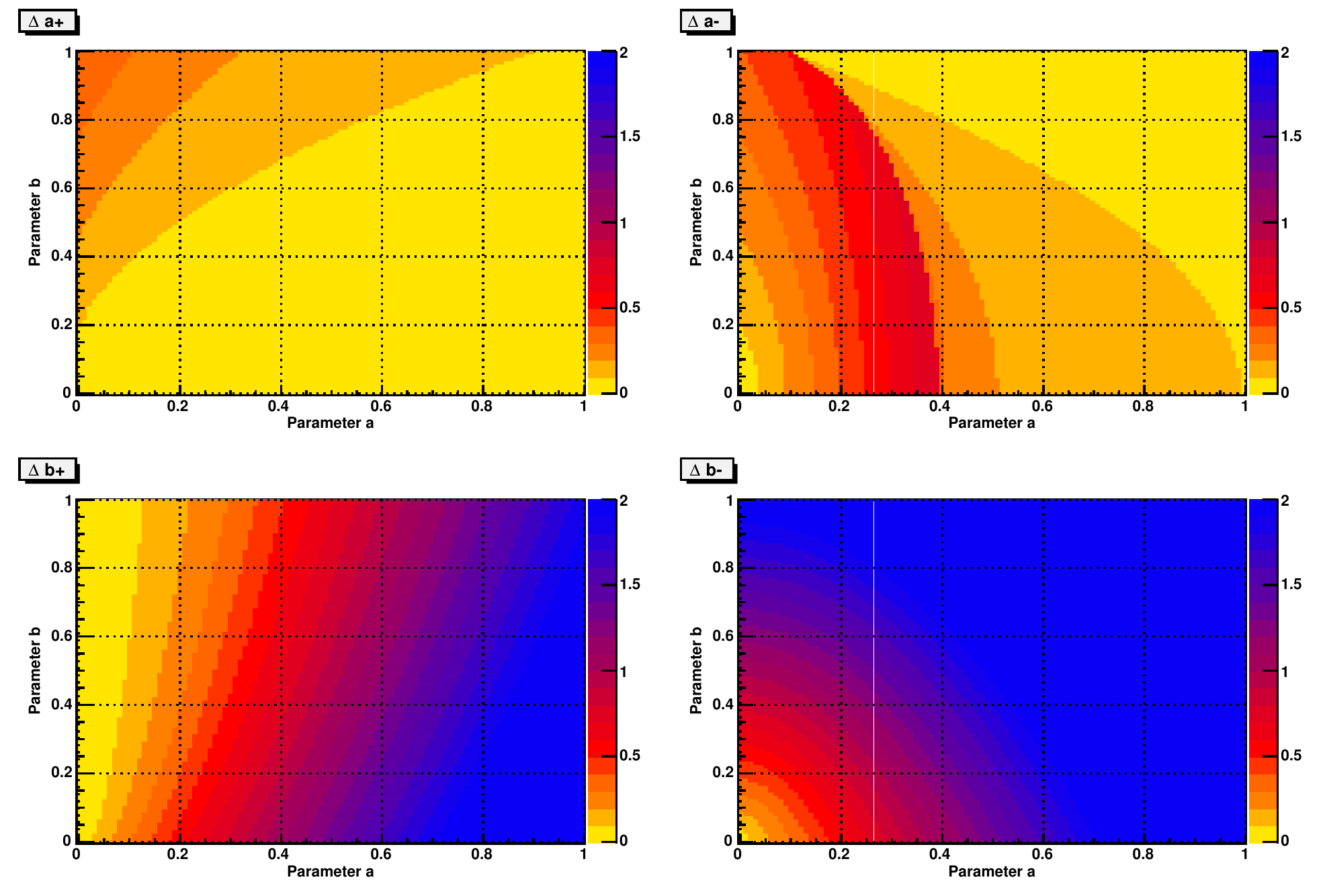}
\caption{\label{fig:sig0sens} Errors $\Delta a^+$ (upper left) and
  $\Delta a^-$ (upper right)  on $a$ as well as $\Delta b^+$ (lower
  left) and $\Delta b^-$ (lower right) on $b$, extracted from the 
  total cross section $\sigma (e^+e^-\to t\bar{t}\Phi)$ at $1 \sigma$
  confidence level for $M_\Phi=120$ GeV at $\sqrt{s}=800$ GeV with
  $\int {\cal L}=500$~fb$^{-1}$. The $e^\pm$ beams are
  unpolarised. The colour code indicates the magnitude of the
  respective error.}
\end{figure}

Figure~\ref{fig:sig0sens} shows the errors on $a,b$ extracted from the
total cross section, and hence the sensitivity of this observable to
the Higgs CP parameters. For the error determination we define a level
of confidence $f$ to identify the area in the $a-b$ plane for which
the value of the observable $O(a,b)$ (here $\sigma (a,b)$) cannot be
distinguished from the reference value $O(a_0,b_0)$ at a specific
point $(a_0,b_0)$. Ignoring systematic errors, it is given by
\beq
|O(a,b)-O(a_0,b_0)| = f \Delta O(a_0,b_0) \;,
\label{eq:sens}
\eeq
where $\Delta O(a_0,b_0)$ is the statistical fluctuation in $O$ at an integrated
luminosity ${\cal L}$ chosen to be 500 fb$^{-1}$ in the following, if
not stated otherwise. For the cross section it is given by
\beq
\Delta \sigma = \sqrt{\frac{\sigma}{{\cal L}}} \;.
\eeq
The errors on $a$ and $b$ for the point $(a_0,b_0)$ are then determined by
the maximal extensions $\Delta a^+$ ($\Delta a^-$) in positive
(negative) $a$ direction and $\Delta b^+$ ($\Delta b^-$) in positive
(negative) $b$ direction, necessary to reach the area
outside the thus defined range of insensitivity. \s  

\begin{figure}[t]
\hspace{0.06\textwidth}
 \begin{minipage}{0.45\textwidth}
       \includegraphics[width=0.78\textwidth]{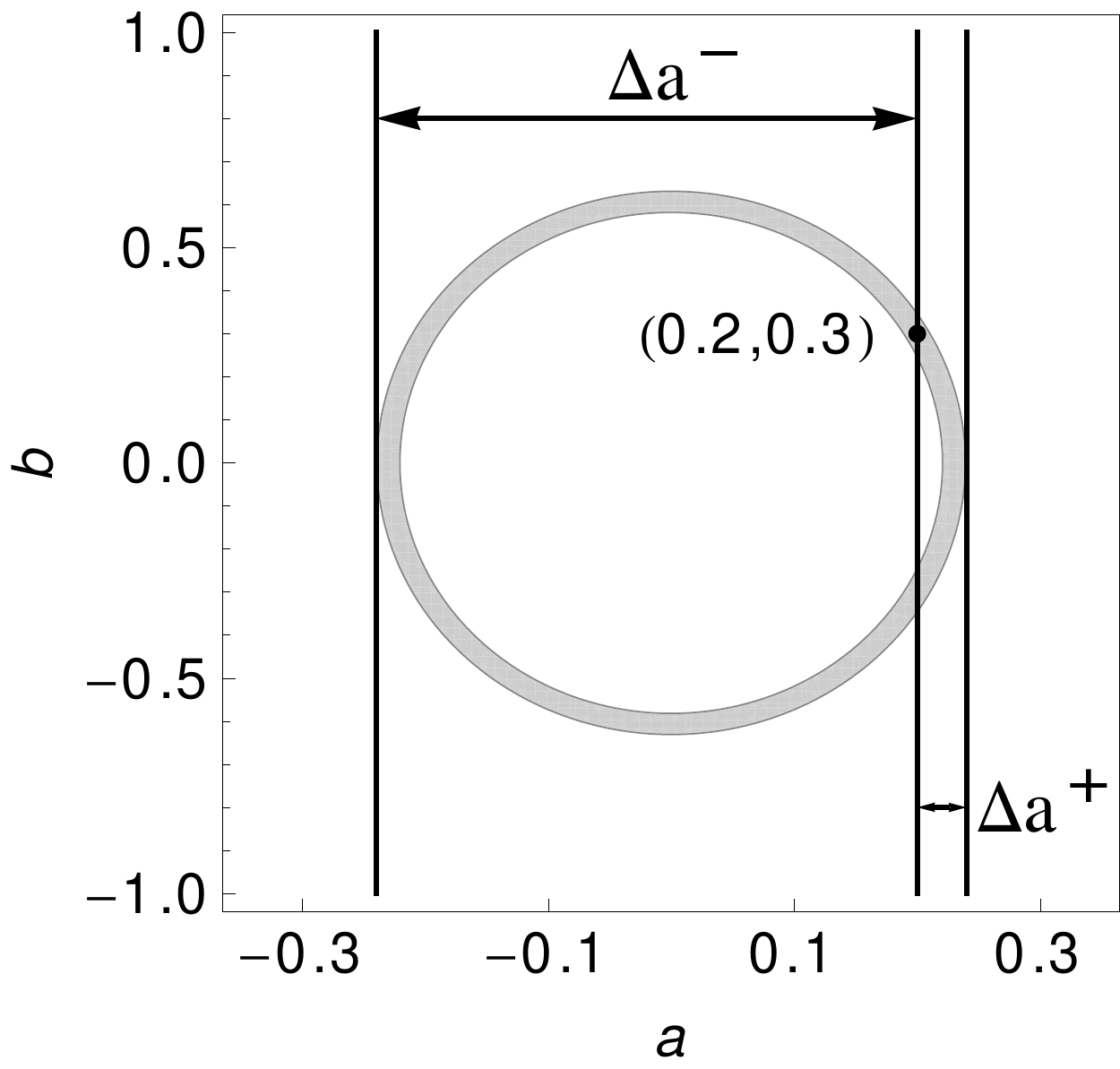}
 \end{minipage}
\hspace{0.01\textwidth}
 \begin{minipage}{0.45\textwidth}
       \includegraphics[width=0.8\textwidth]{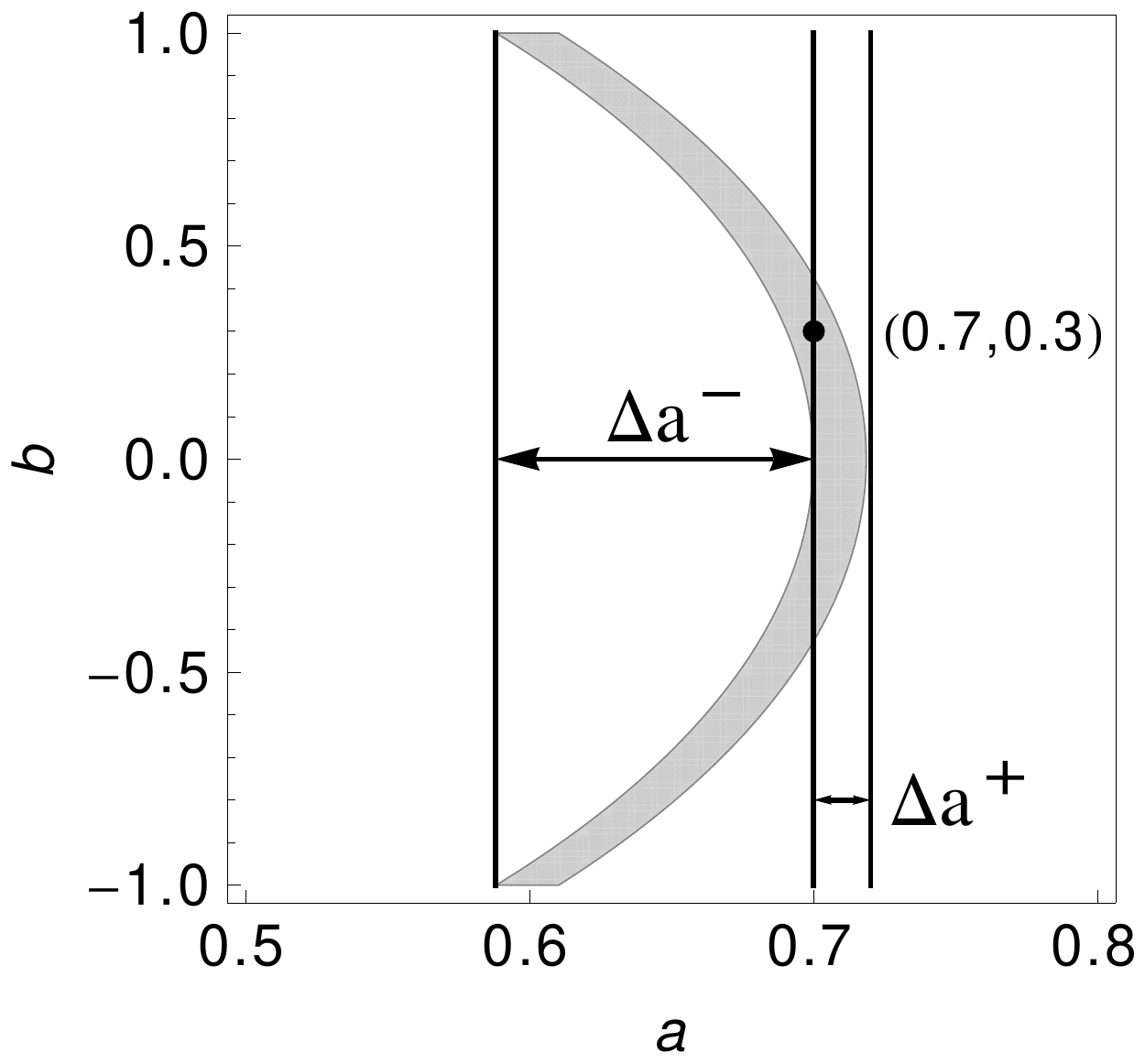}
 \end{minipage}
\hspace{0.03\textwidth}
\caption{\label{fig:errunderstand}  $1\sigma$ band of insensitivity for
  $(a_0,b_0)=(0.2,0.3)$ (left) and $(a_0,b_0)=(0.7,0.3)$ (right). The
  corresponding errors $\Delta a^+$, $\Delta a^-$ are indicated in the
  figures. The errors $\Delta b^{\pm}$ are derived analogously}
\end{figure}

As anticipated from the strong dependence of the cross section on $a$
and as can be inferred from the figure, the cross section is very
sensitive to $a$. In most of the parameter range $a$ can be 
determined with an accuracy of 0.2 or better. 
The error $\Delta a^-$ is large for $a \lsim 0.4$. This can be
understood by realizing that the area of insensitivity at 1$\sigma$
for a point $(a_0,b_0)$ in this parameter region is given
by an elliptic band in the $(a,b)$ plane around this point. Since the
value of $b$ is not known a priori from any other measurement, the
area outside the band of insensitivity is only reached for large
$\Delta a^-$. An illustrative example is shown in
Fig. \ref{fig:errunderstand} (left). For $a\gsim 0.4$ the insensitive
area cuts the edge of the parameter space $|b| \le 1$, so that the
error $\Delta a^-$ is smaller, {\it cf.} Fig.
\ref{fig:errunderstand} (right). The sensitivity to $b$ is only good where
the influence of the dominant contribution to the cross section from
terms proportional to $a$ is small. This is obviously the case for
small $a$ values. The total cross section can therefore be exploited
to determine the CP-even part of a $t\bar{t}\Phi$ coupling. From the
discussion concerning $\Delta a^-$ we expect that the sensitivity on
$a$ will significantly improve if $b$ can be constrained from other
observables. We furthermore found that polarised $e^\pm$ beams improve the
sensitivity only marginally.  

\subsection{The polarisation asymmetry of the top quark}
Since the top decay width is large ($\Gamma_t \sim 1.5$ GeV) the top
quark decays much before hadronization. 
Its spin information is hence translated to the distributions of the
decay products and not contaminated by strong interaction effects.
As the lepton angular distribution in the decay
$t \to bW \to bl \nu$ is not affected by any non-standard effects in
the decay vertex, it is a pure probe of the physics associated with the
top quark production process
\cite{Godbole:2002qu,Godbole:2006tq}. Thus the polarisation asymmetry
$P_t$ of the top quark is another observable expected to probe the
Higgs CP properties. Both for polarised and unpolarised $e^\pm$ beams
it is given by 
\beq
P_t = \frac{\sigma (t_L) - \sigma (t_R)}{\sigma (t_L) + \sigma (t_R)} \;,
\eeq
where $t_{L,R}$ denotes a left-, right-handed top quark. Note that,
since $P_t$ is a CP-even quantity, the polarisation asymmetry of the
anti-top is the same as $P_t$ but with opposite sign. The cross
section for an $L$-,$R$-polarised top quark can be decomposed into a term
$\sigma_0$ proportional to the spin-averaged cross section and a term
$\sigma_1$ proportional to the helicity of the top quark, 
\beq
\sigma _{L,R} = \sigma_0 - \lambda \sigma_1 \;,
\quad \lambda=-1,+1 \quad \mbox{for } L,R \;,
\label{eq:defsig0sig1}
\eeq
with the total cross section given by
\beq
\sigma (t\bar{t}\Phi) = 2 \sigma_0 \;.
\eeq
We can hence express $P_t$ as
\beq
P_t = \frac{2\sigma_1}{2\sigma_0} \; ,
\eeq
where the differential cross section corresponding to $\sigma_1$,
neglecting the small contribution of the diagram involving the
$ZZ\Phi$ vertex, is given by 
\beq
\frac{d\sigma_1}{ dx_1 dx_2} &=& \frac{\alpha^2}{4\pi s} a_t g_{ttH}^2
\left[ \frac{Q_e Q_t v_e}{1-h_z}  + \frac{(a_e^2+v_e^2) v_t}{(1-h_z)^2}\right] 
 \left( a^2 G^H + b^2 G^A \right) \;,
\label{eq:sigma1}
\eeq
with the form factors
\beq
G^H ( x_1,x_2) &=& P (2-x_1-x_2) \Big\{ 
x_1 (1-x_1)  (x_1-x_2) (1-x_2) -16 h_t^2  (2 -x_1-x_2)\nonumber \\
 && -2 h_t [4+x_1 (-3+x_1)  (2+x_1) -6 x_2+8 x_1 x_2
+(1-x_1) x_2^2]+h_\Phi [ x_1(x_2-x_1) \nonumber \\ 
&&  +4 h_t  (2-x_1-x_2)] \Big\}
 \label{eq:G1}
\eeq
and
\beq
G^A (x_1,x_2) &=& P (x_1-x_2) (2-x_1-x_2) \Big\{ h_\Phi (4 h_t+x_1)
+(1-x_1) [ x_1 (1-x_2) \nonumber \\
&& -2 h_t (2-x_1-x_2) ]\Big\} \;.
\label{eq:G2}
\eeq
We have introduced the prefactor
\beq
P = \frac{1}{\sqrt{1-\frac{4h_t}{x_1^2}} (1-x_1)^2 x_1 (1-x_2)^2}
\eeq
and $h_\Phi= M_\Phi^2/s$, $h_t=m_t^2/s$. Since $P_t$ is given by a
ratio of cross sections, this quantity has the advantage of being
independent of an overall model-dependent normalization of the
$t\bar{t} \Phi$ coupling. \s

\begin{figure}[!b]
\vspace*{0.2cm}
 \begin{minipage}{0.5\textwidth}
       \includegraphics[trim = 25mm 15mm 25mm 20mm, clip,width=\textwidth]{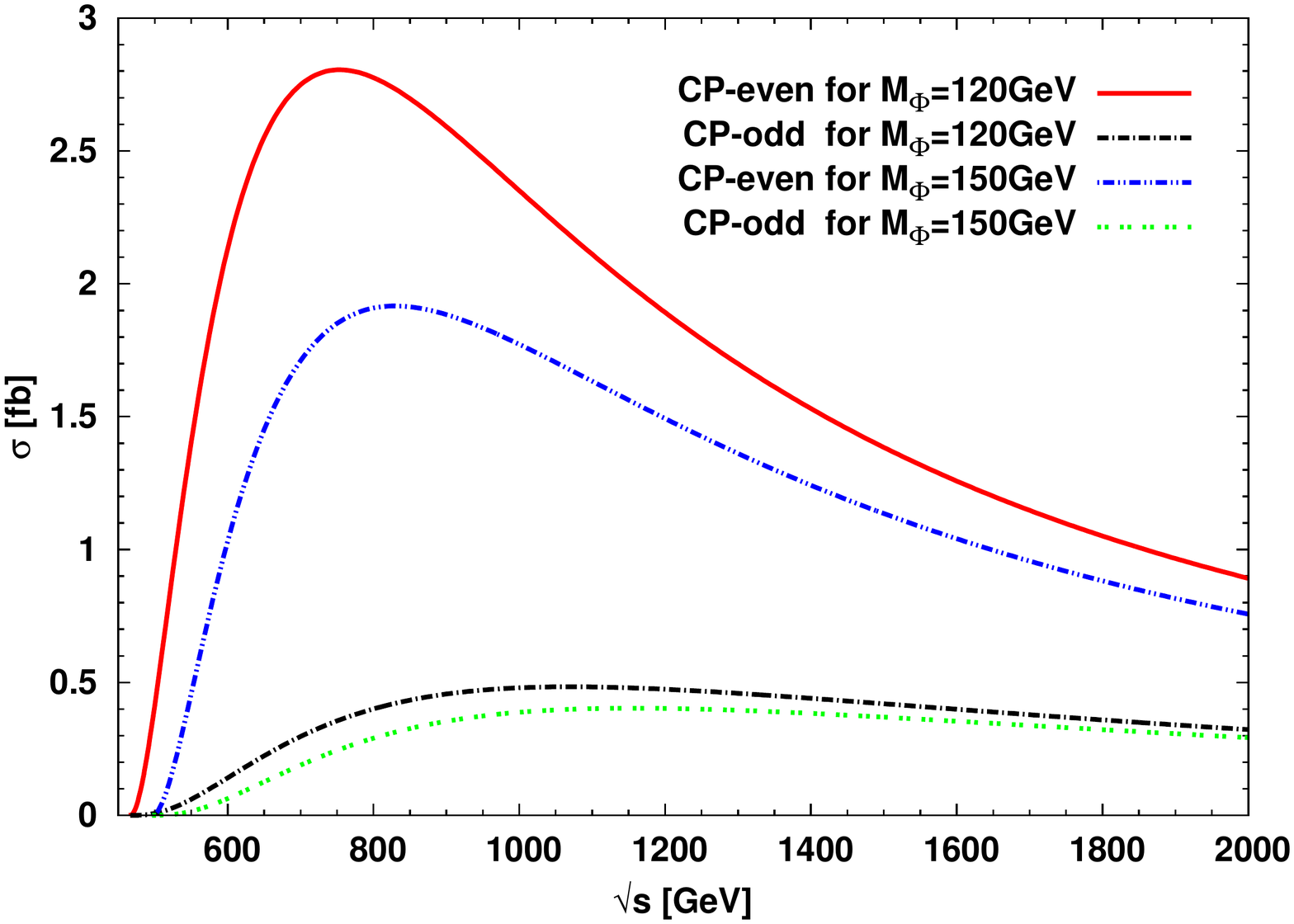}
 \end{minipage}
 \begin{minipage}{0.5\textwidth}
       \includegraphics[trim = 25mm 15mm 25mm 20mm, clip,width=\textwidth]{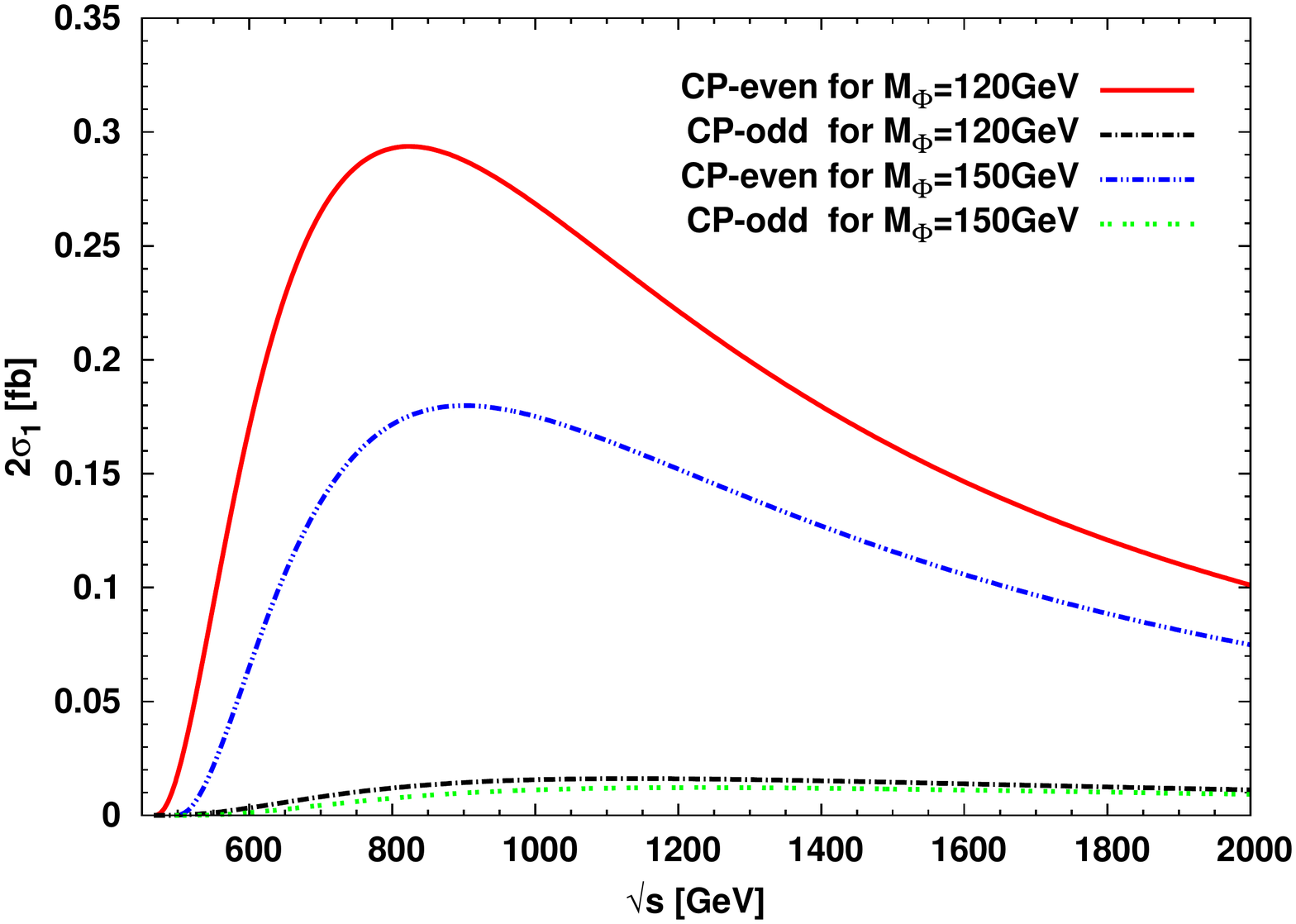}
 \end{minipage}
\caption{\label{fig:cxnrise} The total cross section $\sigma$ (left) and the
  helicity projected cross section $2 \sigma_1$ (right) for a SM
  and a purely pseudoscalar Higgs boson as a function of $\sqrt{s}$ for two
  masses $M_\Phi=120$ and 150 GeV with unpolarised $e^\pm$ beams.}
\vspace*{0.4cm}
\end{figure}

\begin{figure}[t]
\vspace*{-0.8cm}
\centering 
\includegraphics[width=10cm]{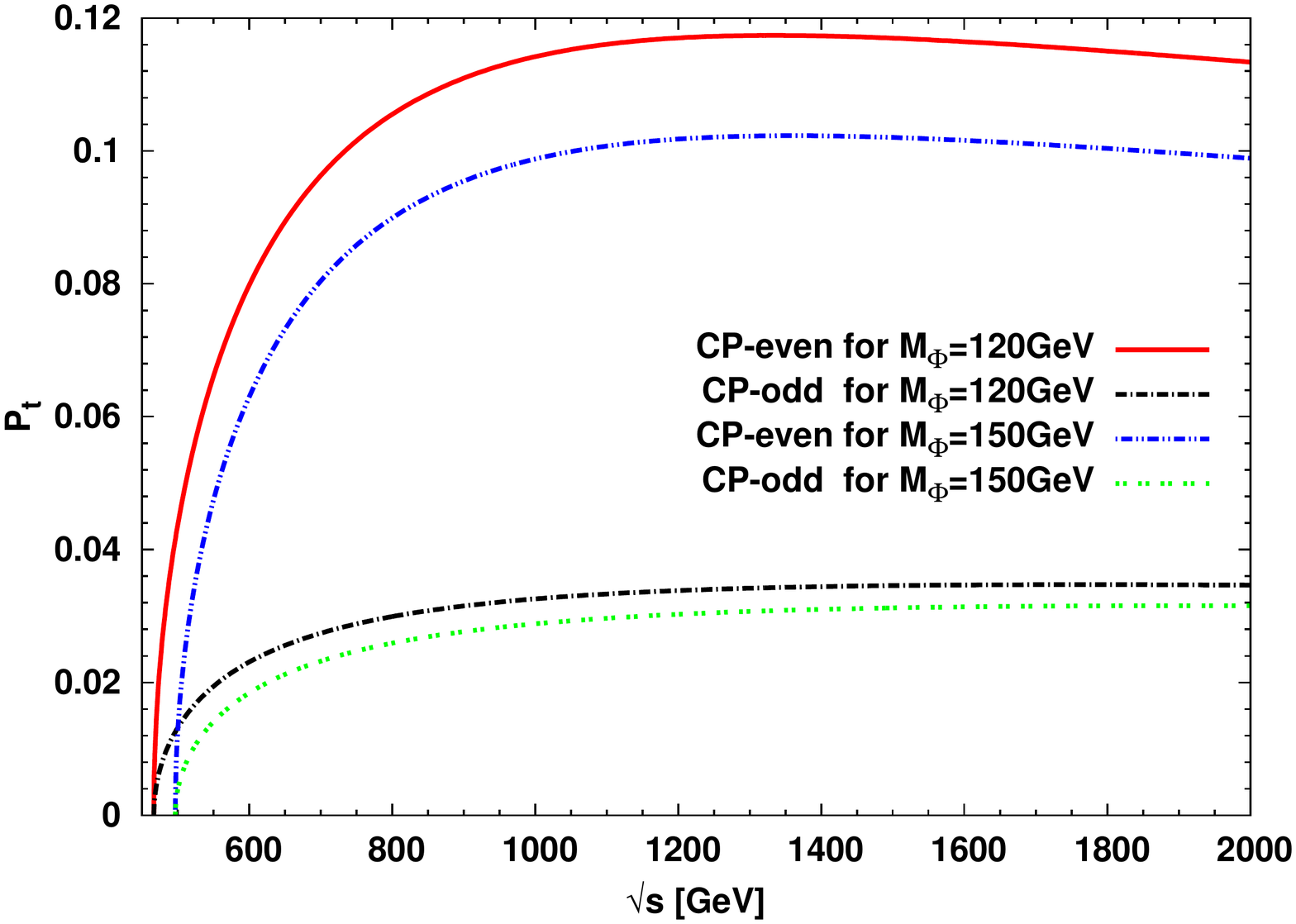}
\vspace*{-0.6cm}
\caption{\label{fig:ptrise} The top polarisation asymmetry for a SM and a
  purely pseudoscalar Higgs  boson as a function of $\sqrt{s}$ for two masses
  $M_\Phi =120$ and 150 GeV with unpolarised $e^\pm$ beams.}
\end{figure}

The total cross section $\sigma$
and the helicity projected cross section $2\sigma_1$  are shown in
Fig.~\ref{fig:cxnrise} (left) and (right), respectively, as a function of
$\sqrt{s}$ for a scalar SM Higgs and a purely pseudoscalar Higgs
and for two different Higgs masses $M_\Phi = 120$ and 150 GeV. 
Like in the case of the total cross section, $\sigma_1$  shows a flat threshold
rise for the pseudoscalar and a steep one for the scalar Higgs. The
dependence on $\rho$, which parametrizes the small deviation from the
threshold c.m. energy, differs by approximately one power for the
CP-even and the CP-odd case, like in the case of $\sigma =
2\sigma_0$. This results in a threshold rise for $P_t$ 
which is approximately the same for the scalar and the pseudoscalar
Higgs as can be inferred from Fig.~\ref{fig:ptrise}. Due to the
reduced phase space the cross section and the top polarisation
asymmetry decrease with increasing Higgs mass. \s 

Furthermore, the values of $\sigma_0$, $\sigma_1$ and of $P_t$
are very different depending on the CP quantum numbers of the spin 0
state. The difference is solely due to different form factors
$F_{1,2}^H,F_{1,2}^A$ ($G^H,G^A$) for the scalar and pseudoscalar
Higgs in $\sigma_0$ ($\sigma_1$). There is no difference in the
coupling structure for the scalar $H$ and the pseudoscalar $A$, {\it
  cf.} Eqs.~(\ref{ttHxsection}) and (\ref{eq:sigma1}). At very high
energies, the chiral limit is reached and the form factors for $H$ and
$A$ (and hence the corresponding values of $\sigma_0$ and $\sigma_1$)
become equal for a scalar and a pseudoscalar Higgs, up to the contributions
from the diagram with the $ZZ\Phi$ coupling, which are subleading at
high c.m. energy. This behaviour can be anticipated from
Figs.~\ref{fig:cxnrise} (left) and (right). For the polarisation
asymmetry, being the ratio of $\sigma_0$ and $\sigma_1$,  the chiral
limit is reached much more slowly. We also studied this behaviour of
$P_t$ at high energies in the process $e^+e^- \to b\bar{b}\Phi$ which,
apart from the couplings and masses, is identical to $t\bar{t} \Phi$
production. Due to the smaller final state masses, however, the
process takes place at lower energies and the chiral limit can be
observed at much lower energies compared to the $t\bar{t}\Phi$ final state. 
\s

\begin{figure}[t]
\vspace*{-0cm}
\begin{center}
\includegraphics[width=9cm]{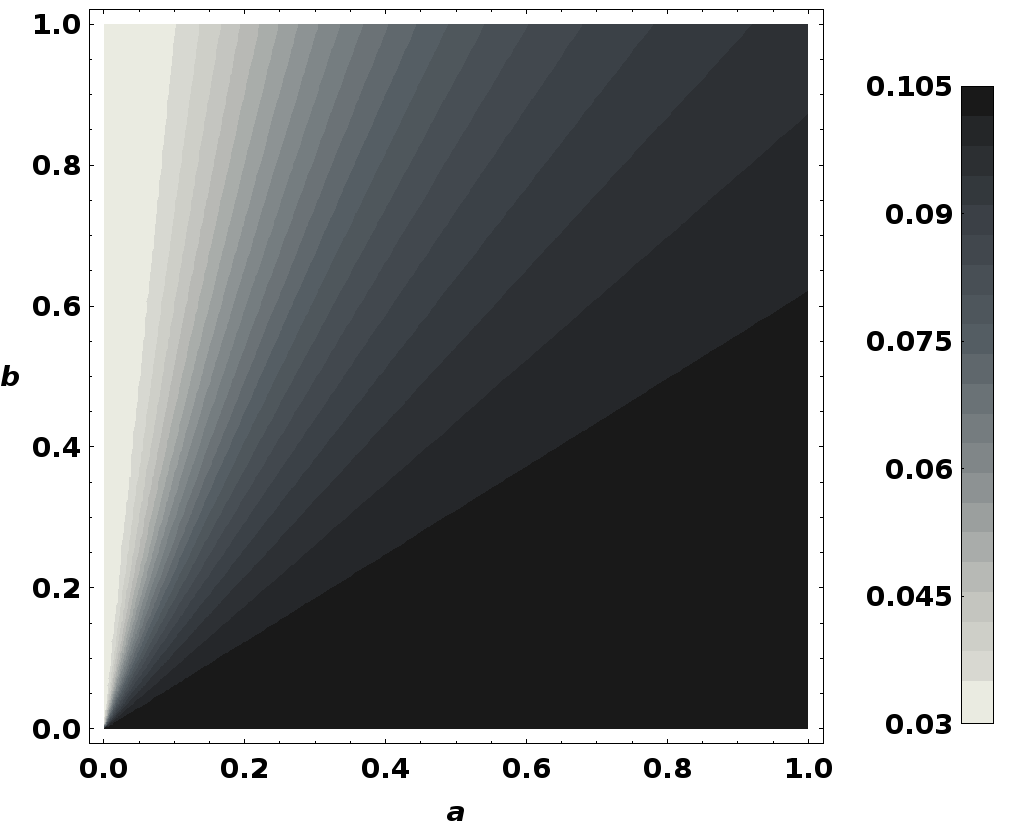}
\caption{\label{fig:ptcontour} Contour plot of the top
  polarisation asymmetry $P_t$ in the $a-b$ parameter plane for 
  $M_H=120$ GeV and $\sqrt{s}=800$ GeV. The $e^\pm$ beams are
  unpolarised. The grey code indicates the value of $P_t$.}
\end{center}
\end{figure}

The values of $P_t$ are shown as contour plots in the $a-b$
parameter plane in Fig.~\ref{fig:ptcontour} for $M_\Phi=120$ GeV,
$\sqrt{s}=800$ GeV and unpolarised $e^\pm$ beams. 
The top polarisation asymmetry in case of a purely pseudoscalar state is with 
0.03 much smaller than the one for a SM Higgs boson with 0.11.
Furthermore, $P_t$ rapidly increases with $a$ for large values of
$b$ and small values of $a$ and then approaches a nearly constant behaviour
in $a$. For small $b$, $P_t$ is constant in $a$ (due to our choice
$c=-a$), as the $b$ contribution to $\sigma_0$ and $\sigma_1$
does not play a role and can be neglected. The coupling factor $a^2$
hence cancels in the ratio of $\sigma_1$ and $\sigma_0$ so that $P_t$
is simply given by the corresponding ratio of the form factors and is
independent of $a$. The same of course holds for small values of
$a$. Here $P_t$ is almost constant in $b$. Overall, the change of
$P_t$ with $b$ is small. 
The largest values of $P_t$ are reached for large $a$. Polarising
the $e^\pm$ beams increases $P_t$ independently of the CP nature of
the Higgs boson by roughly a factor of 3. The replacements to be made in
the formulae of $\sigma_0,\sigma_1$ for polarised beams are given in
the Appendix.\s  

The errors on the extraction of $a$ and $b$ from the top polarisation
asymmetry have been calculated as well, with $\Delta O(a_0,b_0)$ of
Eq.~(\ref{eq:sens}) given by 
\beq
\Delta O(a_0,b_0) = \Delta P_t = \frac{1}{\sqrt{\sigma {\cal L}}}
\sqrt{1-P_t^2} \;. 
\eeq 
The errors on $a$ and $b$ for unpolarised beams turn out to be larger than 1.
Polarisation only slightly improves the situation. Whereas the observable
$P_t$ can be used to distinguish a pseudoscalar from a scalar Higgs,
the use of $P_t$ {\it alone} does not allow for an accurate extraction of
$a$ and $b$, unless the parameter range has been constrained elsewhere
before. 

\subsection{Ratios of cross sections}
Scalar and pseudoscalar Higgs production differ by
their threshold behaviour. A simple method to test this different
behaviour is the measurement of the total cross section at two
different c.m. energies. These should both be below $\sim 800$ GeV, so
that they lie within the region of rising $\sigma$, and the
number of events should not be too small. As c.m. energies we
therefore chose $\sqrt{s_1}=800$ GeV and $\sqrt{s_2}=600$ GeV and
investigated the ratio 
\beq
R = \frac{\sigma (\sqrt{s_1})}{\sigma (\sqrt{s_2})} \;.
\eeq
The advantage of investigating ratios of cross sections is that the
effect an overall model-dependent normalization of the top
quark Yukawa coupling and some systematic errors in 
the measurement can be avoided.  We found, however, that the results for
$\Delta a^\pm$ and $\Delta b^\pm$ show the same behaviour as for
the observable $P_t$. The ratio $R$ is a good observable to
distinguish a purely CP-odd from a CP-even Higgs boson in case of a
CP-conserving Higgs sector, or to distinguish a CP-mixed Higgs state
with a large CP-odd contribution from a CP-even Higgs. It is
difficult, however, to extract the absolute values of $a$ and $b$ in a
model-independent approach over the whole considered $a-b$ parameter
range. In the following we will therefore not use $R$, but instead
$P_t$ (together with the total cross section and the up-down
asymmetry) to determine the CP quantum numbers. 
It may be possible to measure the ratio $R$ more accurately, but
this requires running at two different collider energies, whereas for
the measurement of $P_t$, while more complex, running at a single
energy suffices. 

\subsection{The CP-violating up-down asymmetry}
\begin{figure}[t]
\vspace*{-0cm}
\begin{center}
\includegraphics[width=9cm]{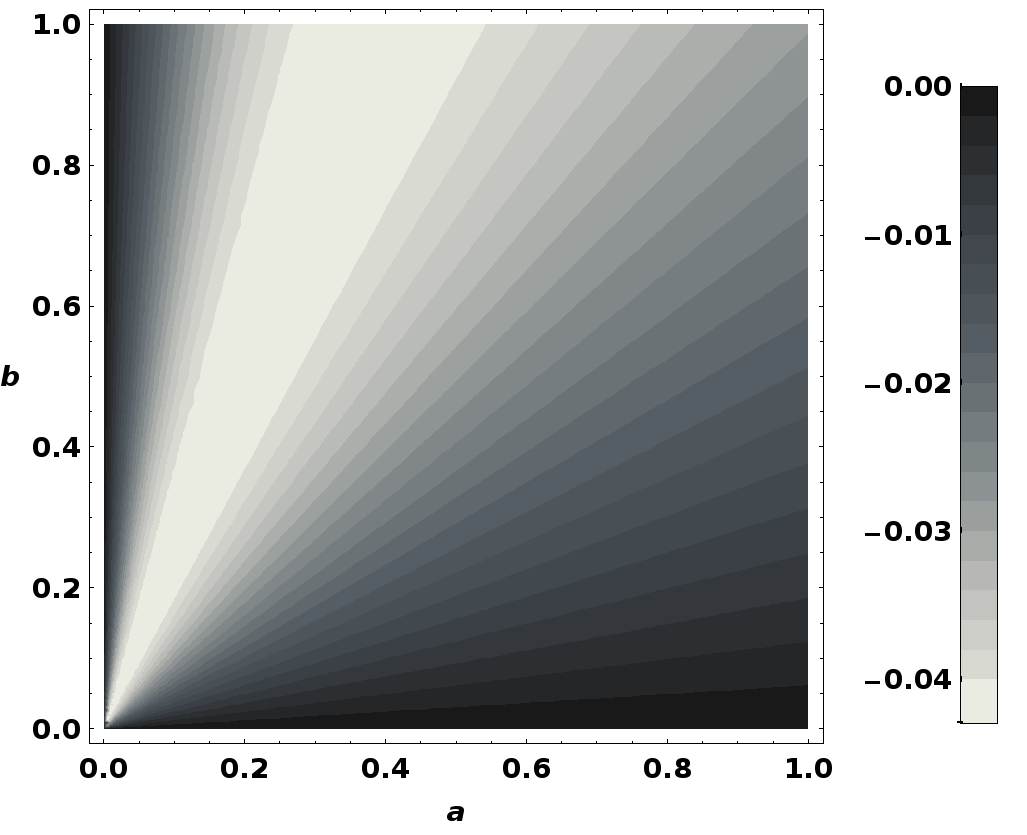}
\caption{\label{fig:updownunpol} Contour plot of the up-down asymmetry
  $A_\phi$ in the $a-b$ parameter plane for $M_H=120$ GeV and
  $\sqrt{s}=800$ GeV. The $e^\pm$ beams are unpolarised. The grey
  code indicates the value of $A_\phi$.}
\end{center}
\vspace*{-0.2cm}
\end{figure}

Contrary to the total cross section and the top polarisation asymmetry
the up-down asymmetry $A_\phi$ of the antitop with respect to the
top-electron plane is an observable which is sensitive to CP violation. 
Defining by $q_{a,b}$ the four-momenta of the incoming $e^-,e^+$ and
by $p_{1,2}$ the four-momenta of the top, antitop,
respectively, the angle $\phi$ between the antitop direction and
the top-electron plane is given by
\beq
\sin \phi = \frac{\vec{p}_2 (\vec{q}_a \times
  \vec{p}_1)}{|\vec{p}_2||\vec{q}_a \times \vec{p}_1|} \sim
\epsilon^{p_1 p_2 q_a q_b} \;,
\eeq
with the totally antisymmetric Levi-Civita tensor $\epsilon$.
The up-down asymmetry of the $t\bar{t}\Phi$ cross section $\sigma$ is
defined as 
\beq
A_\phi = \frac{\sigma (\mbox{up}) - \sigma(\mbox{down})}{\sigma
  (\mbox{up}) + \sigma(\mbox{down})} \;,
\eeq
where 'up'  ('down') denotes the value of the cross section if the integration
over $\phi$ is performed for  $\phi \in [0,\pi)$ ($\phi \in
[\pi,2\pi)$). It turns out that $A_\phi$ is given by the interference of 
the diagram where the Higgs is radiated from the top-quark with the
diagram where the Higgs is radiated from the $Z$ boson
\cite{BarShalom:1995jb}. The asymmetry can be cast
into the form 
\beq
A_\phi = \frac{\sigma_{as} bc}{\sigma} \;,
\label{eq:aphi}
\eeq
with the numerator of Eq.~(\ref{eq:aphi}) denoting the asymmetric
interference term proportional to $bc$, which can be derived from the
differential cross section given in Eq.~(\ref{eq:updownas}) of the Appendix. The
denominator is given by the total cross section $\sigma=\sigma
  (\mbox{up}) + \sigma(\mbox{down})$. A non-vanishing $A_\phi$ is
an unambiguous indicator of CP violation. As it is proportional to $b$,
the $t\bar{t} \Phi$ coupling must contain a CP-odd part. If, however, the Higgs
sector is CP-conserving and a purely CP-odd Higgs boson is radiated
from the top line, the coupling to the $Z$ boson must vanish at
tree-level, {\it i.e.} $c=0$, so that in the CP-conserving case $A_\phi=0$.
The contour lines of $A_\phi$ are shown in Fig. \ref{fig:updownunpol}
in the $a-b$ parameter plane for $M_\Phi = 120$ GeV and $\sqrt{s}=
800$ GeV. The incoming $e^\pm$ beams are unpolarised. 
As we put $c=-a$, the up-down asymmetry depends in the numerator
linearly on $a$ and $b$. In the denominator it depends
via the total cross section quadratically on $a$ and $b$. The
asymmetry is shown for positive values $a$ and $b$. It can be inferred
for the negative values from its asymmetric behaviour with respect to
a sign change of $a$ or $b$. We found that the polarisation of the
initial beams does not help to increase $A_\phi$. The polarisation of
the $e^\pm$ beams contributes differently to the various terms
entering the total cross section, which is in the denominator of
$A_\phi$, and to the interference term $\sim bc$ in the numerator of
$A_\phi$. This results in a small net effect on $A_\phi$ of the initial beam
polarisation. This can also be inferred from the formulae of the
various terms for polarised $e^\pm$ beams which can be derived by
using the formulae given in the Appendix. For
higher c.m. energies the asymmetry is larger, as the cross section in
the denominator decreases faster with rising $\sqrt{s}$ as the
numerator. However, the total number of events will decrease for the
same reason, which can be balanced by a higher luminosity. \s  

The errors on $a$, $b$ have been extracted in the same way as before
with $\Delta O(a_0,b_0)$ of Eq.~(\ref{eq:sens}) given by 
\beq
\Delta O(a_0,b_0) = \Delta A_\phi = \frac{1}{\sqrt{\sigma {\cal L}}}
\sqrt{1-A_\phi^2} \;. 
\eeq 
The errors extracted solely from $A_\phi$ turn out to
be larger than the absolute values of $a,b$. This is the result of an interplay
of a too small variation of $A_\phi$ with $a$ and/or $b$ and too large
errors $\Delta A_\phi$. \s

\begin{figure}[t]
\vspace*{-1cm}
\centering 
\includegraphics[width=12cm]{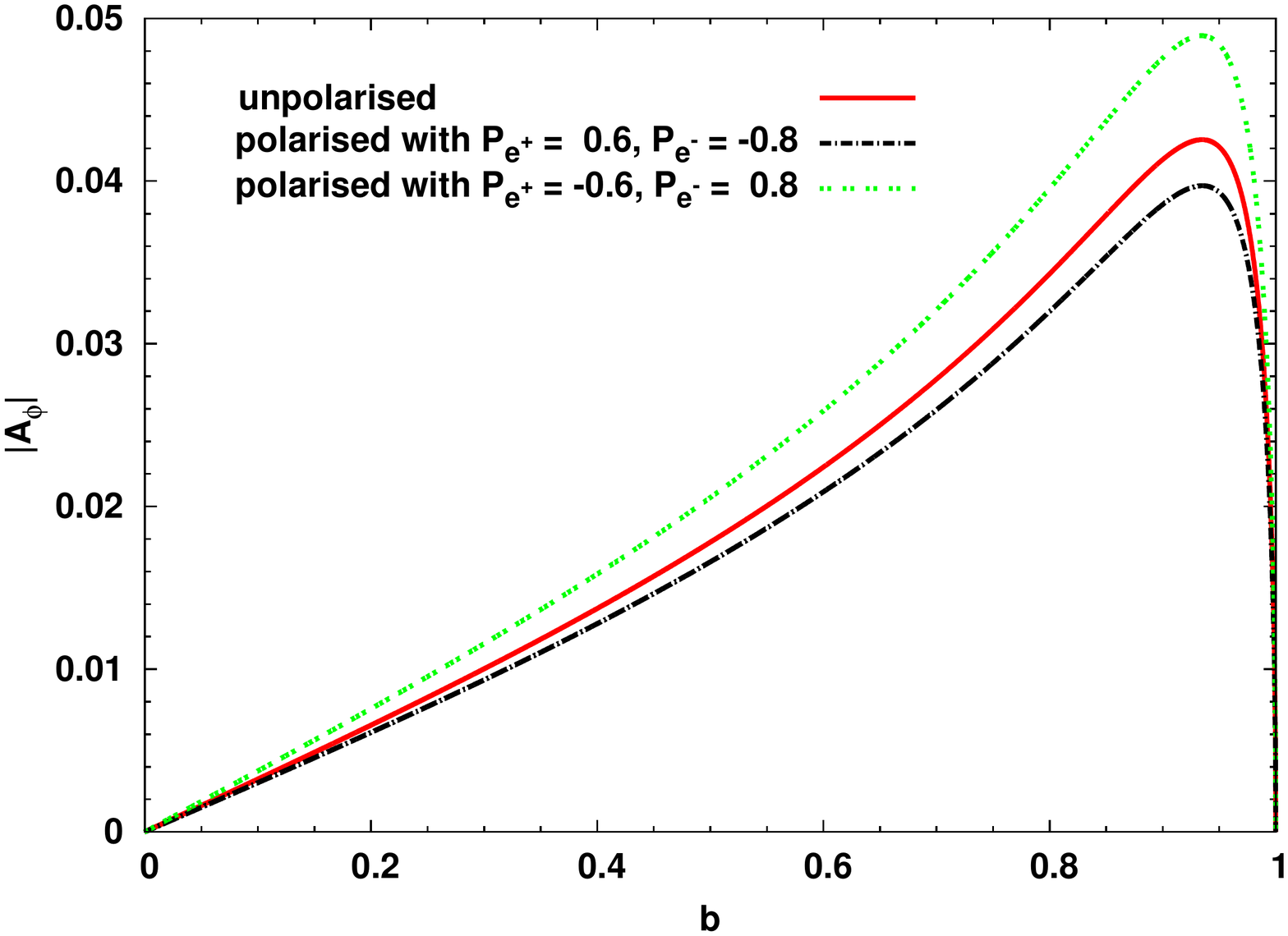}
\vspace*{-0.6cm}
\caption{\label{fig:aphifuncb} The absolute value of the up-down
  asymmetry $A_\phi$ as function of $b$ with $a^2+b^2=1$, for
  unpolarised beams (red/full), polarised beams with
  $P_{e^-}=-0.8,P_{e^+}=0.6$ (black/dash-dotted) and polarised beams
  with $P_{e^-}=0.8,P_{e^+}=-0.6$ (green/dotted).}
\end{figure}

We conclude this subsection by showing in Fig. \ref{fig:aphifuncb} the
up-down asymmetry in case we impose $a^2=1-b^2$. Being 
proportional to $b$ it vanishes for $b=0$. As it is also proportional to
$c$, at $b=1$, where $a=0$, it vanishes due to our choice $c=-a$. As
can be inferred from the figure, switching the polarisation from
$P_{e^-}=-0.8,P_{e^+}=0.6$ to $P_{e^-}=0.8,P_{e^+}=-0.6$ slightly
increases $A_\phi$. 
In this case, however, the cross section is not increased by
about a factor of 2 any more as in the case of our original choice of 
$P_{e^\pm}$, but is decreased by about 10\%.

\subsection{The combined sensitivity}
\begin{figure}[h]
\vspace*{-0.8cm}
\hspace*{-0.3cm}
\begin{center}
\includegraphics[width=16cm]{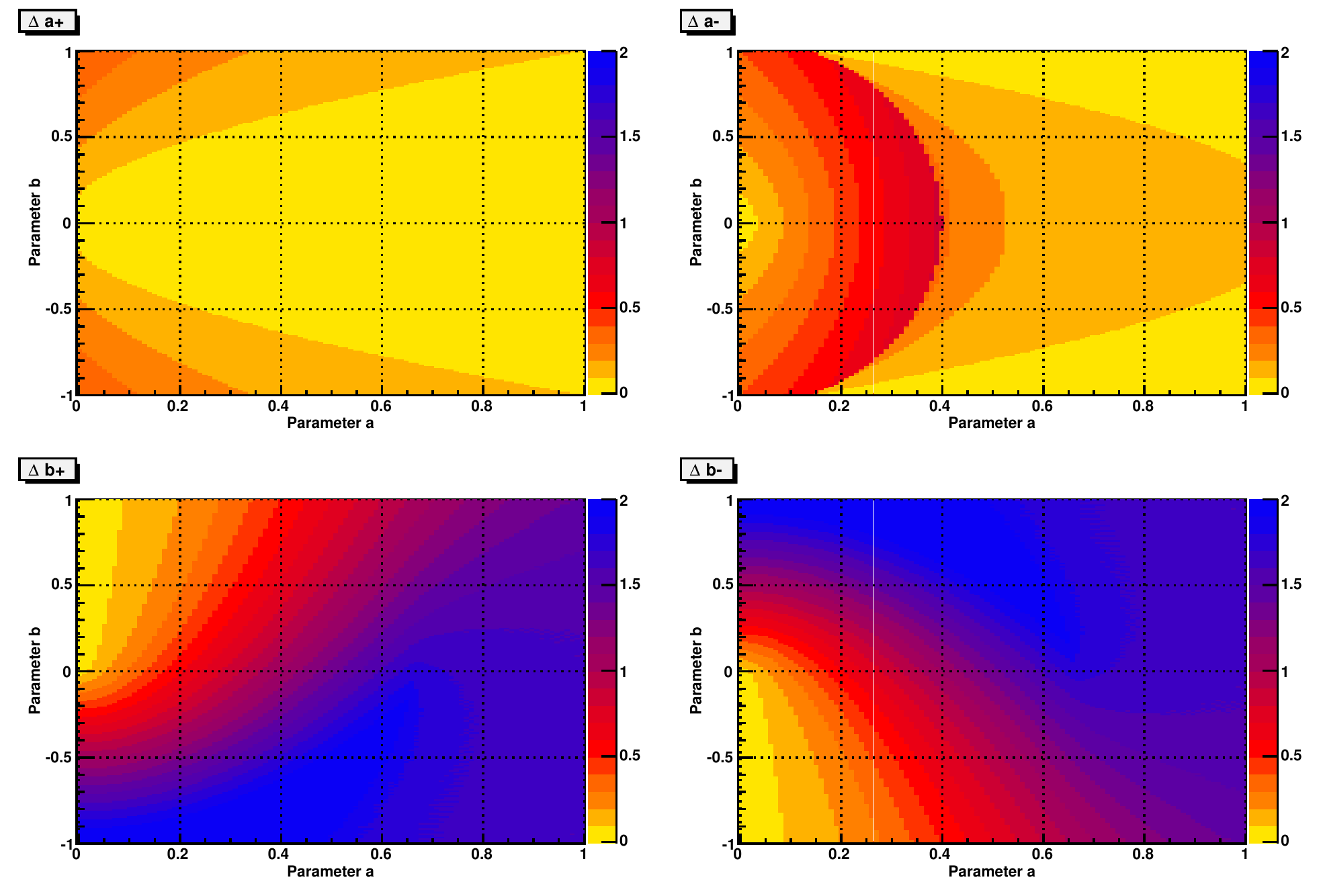} 
\caption{\label{fig:allsens} Errors $\Delta a^+$ (upper left) and
  $\Delta a^-$ (upper right)  on $a$ as well as $\Delta b^+$ (lower
  left) and $\Delta b^-$ (lower right) on $b$, by combining all 3
  observables $\sigma, P_t, A_\phi$, at $1 \sigma$ confidence level
  for $M_\Phi=120$ GeV and $\sqrt{s}=800$ GeV with ${\cal  L} = 500$
  fb$^{-1}$. The $e^\pm$ beams are unpolarised. The colour code
  indicates the magnitude of the respective error.} 
\end{center}
\end{figure}
\begin{figure}[t]
\vspace*{-0.8cm}
\hspace*{-0.3cm}
\begin{center}
\includegraphics[width=16cm]{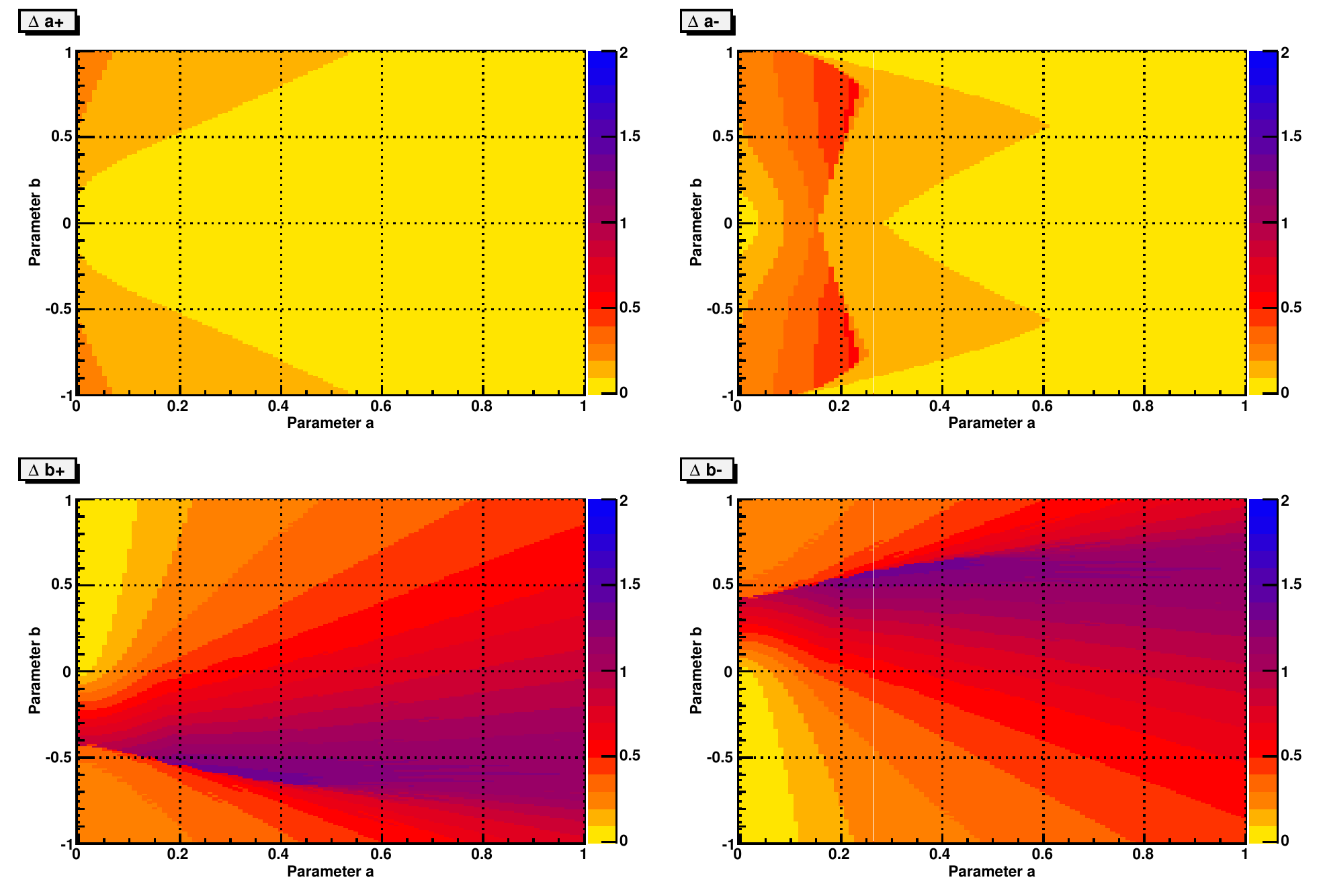}
\caption{\label{fig:allpolsens} Same as Fig. \ref{fig:allsens}, but for
  polarised $e^\pm$ beams.} 
\end{center}
\end{figure}

 
As has been discussed in the previous subsections the total cross
section is a good observable to determine $a$, whereas less good for
the measurement of $b$. The top polarisation asymmetry, on the other
hand, can only be used to distinguish a CP-odd from a CP-even Higgs
boson. The up-down asymmetry $A_\phi$ tests CP-mixing, but 
the errors on $a$ and $b$ are too large to be useful for a determination
of $a,b$. The accuracy on $a$ ($b$) will substantially improve,
however, if the parameter $b$ ($a$) has been extracted beforehand from
some other measurement. We therefore combine all three observables
$O_i$ ($i=1,2,3$) given by $\sigma$, $P_t$ and $A_\phi$ to derive new
sensitivity areas for $a$ and $b$ by performing a $\chi^2$ test, with 
$\chi^2$ defined as
\beq
\chi^2 = \sum_{i=1,2,3}\frac{(O_i (a,b)-O_i (a_0,b_0))^2}{(\Delta O_i
  (a_0,b_0))^2}  \;.
\eeq
The errors $\Delta a^\pm, \Delta b^\pm$ are again determined by the maximal
extensions of the insensitive areas and are shown at $1 \sigma$
confidence level in Fig.~\ref{fig:allsens} for unpolarised and in
Fig.~\ref{fig:allpolsens} for polarised beams with $\sqrt{s}=800$ GeV
and $\int {\cal L}=500$ fb$^{-1}$.  \s

In case of unpolarised beams the plots essentially reproduce the
errors on $a$ and $b$ extracted from the total cross section
alone. This was expected since it is the most sensitive observable and
$P_t$ and $A_\phi$ do not help to improve the errors in this case.
If the initial beams are polarised, however, the errors on $a$ are
remarkably reduced. This is due to the
mutual interplay between $\sigma$ and $P_t$ in constraining the
parameter ranges of $a$ and $b$. The polarisation is necessary in
order to increase the cross section and $P_t$. For $P_t$ this also leads
to a smaller statistical fluctuation $\Delta P_t \sim 1/\sqrt{\sigma
  {\cal L}}$. As for $b$, polarisation slightly helps to decrease the
error for $|b| \gsim 0.5$.  The observable $A_\phi$ does not have any
effect on the reduction of the errors at $\sqrt{s}=800$ GeV. \s

\begin{figure}[th]
\vspace*{-0.8cm}
\hspace*{-0.3cm}
\begin{center}
\includegraphics[width=16cm]{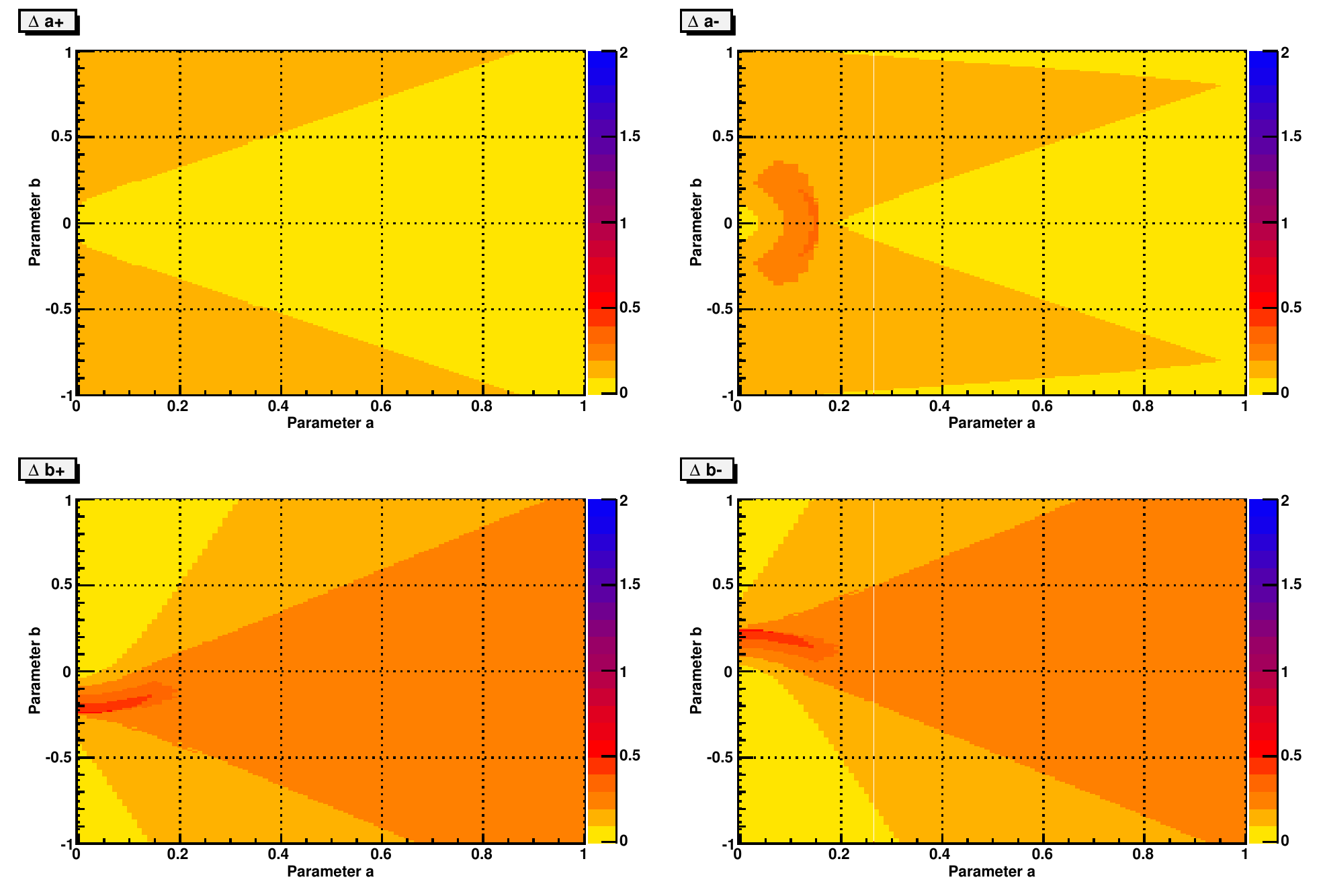}
\caption{\label{fig:allsens3000pol} Errors $\Delta a^+$ (upper left) and
  $\Delta a^-$ (upper right)  on $a$ as well as $\Delta b^+$ (lower
  left) and $\Delta b^-$ (lower right) on $b$, by combining all 3
  observables $\sigma, P_t, A_\phi$, at $1 \sigma$ confidence level
  for $M_\Phi=120$ GeV and $\sqrt{s}=3$ TeV with ${\cal  L} = 3$
  ab$^{-1}$. The $e^\pm$ beams are polarised. The colour code
  indicates the magnitude of the respective error.} 
\end{center}
\end{figure}

At high c.m. energies of  $\sqrt{s} = 3-5$ TeV as realized in the CLIC
design \cite{Accomando:2004sz} the total cross section, which scales
with the energy, is smaller. On the other hand this will be balanced
by higher integrated luminosities of $\int {\cal L} = 3-5$
ab$^{-1}$, so that the signal rate approximately remains the same. 
The top polarisation asymmetry does not change a lot
compared to 800 GeV c.m. energy, as can be extrapolated from
Fig.~\ref{fig:ptrise}, which shows $P_t$ as a function of $\sqrt{s}$. 
The statistical fluctuation $\Delta P_t$, which is inversely
proportional to the signal rate, almost does not change either. The
up-down asymmetry, however, gets larger with rising $\sqrt{s}$. 
Therefore at high energies in the multi-TeV regime all three
observables contribute significantly to $\chi^2$. Their mutual
interplay results in remarkably small errors on $a$ less than about
0.2 in large parts of the parameter space. This can be inferred from
Fig. \ref{fig:allsens3000pol}, which shows for $\sqrt{s} = 3$ TeV and
${\cal L} = 3$ ab$^{-1}$ the errors on $a$ and $b$ at 1$\sigma$
confidence level extracted from the combination of all three observables 
for polarised beams. It turns out that $P_t$ and $A_\phi$ contribute
to the reduction of the error in complementary areas of the $a-b$
parameter plane. As for $b$, the error is $\lsim 0.3$ except for
low values of $a$ and $b$ values around $\sim \pm 0.2$, where it gets
worse. \s  

\section{Radiation of a spin 1 particle}
In this section we investigate the question to what extent the spin
of the particle produced in association with a $t\bar{t}$ pair
affects the total cross section and the polarisation asymmetry. The
motivation is two-fold. On the one hand $t\bar{t}Z$ production
contributes a large part to the irreducible SM background of
associated Higgs production, so that a distinction of these two
processes for the identification of the signal process is necessary
\cite{Juste:1999af}. On the other hand numerous models of New Physics
predict the existence of additional neutral particles $Z'$ with spin
$J=1$. In the following the associated production of such a particle
will be investigated without referring to a specific
model.\footnote{See Ref.~\cite{Baur:2004uw} for associated SM $Z$ 
  boson production with anomalous couplings.}
This particle is only demanded to couple to top quarks but not to
$e^\pm$, like the SM Higgs boson.\footnote{A non-vanishing coupling
  $eeZ'$ would have been excluded by the LEP experiments for the
  masses we study.} 
The description of the $t\bar{t} Z'$ coupling in an effective
model-independent ansatz demanding Lorentz invariance and hermiticity
and including terms of up to dimension 5 depends on ten parameters
\cite{Hollik:1998vz}.  
In a first approach we do not take into account anomalous vector couplings
to $t\bar{t}$ of dimension higher than 4. This will be left to a
future publication. The $t\bar{t}Z'$ coupling is hence given by
\beq
C_{ttZ'} = -ie \gamma_\mu (g_V - g_A \gamma_5) \;,
\eeq
with the vector and axial-vector couplings $g_V$ and $g_A$
chosen to be of ${\cal O}(1)$.\s

\begin{figure}[t]
\vspace*{-0.8cm}
\hspace*{-0.3cm}
\begin{center}
\includegraphics[trim = 0mm 0mm 0mm 0mm, clip,width=16cm]{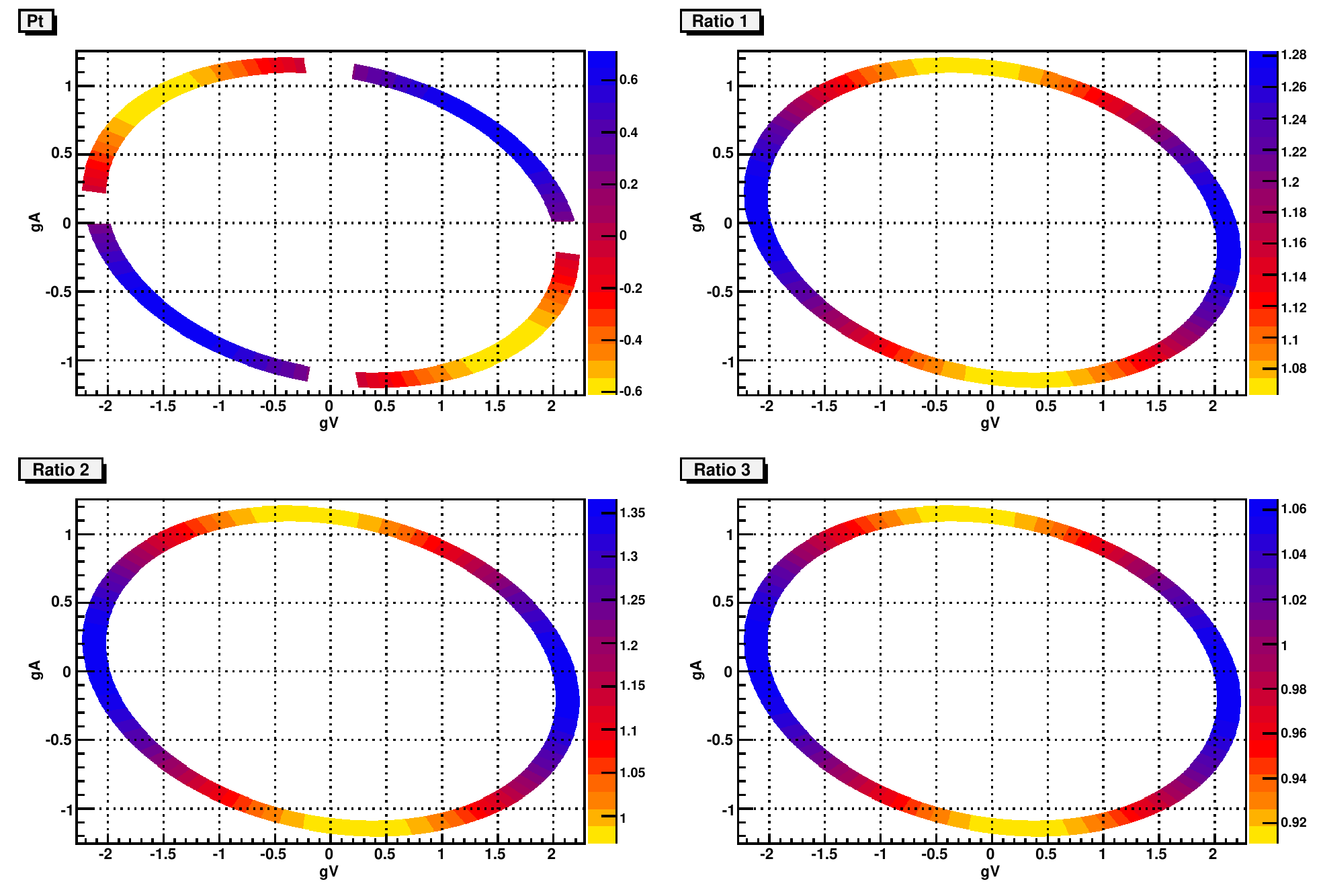}
\caption{\label{fig:ptspin1} The $P_t$ contours for
  $t\bar{t}Z'$ production with $M_{Z'}=120$ GeV and $\sqrt{s}=800$ GeV
  (upper left). The contours for the ratio of the cross sections at
  different c.m. energies are shown in the upper right plot for 
  $\sigma (1000 \mbox{ GeV})/\sigma(800 \mbox{ GeV})$, the lower left
  plot for $\sigma (1300 \mbox{ GeV})/\sigma(800 \mbox{ GeV})$ and the
  lower right plot for $\sigma (1300 \mbox{ GeV})/\sigma(1000 \mbox{
    GeV})$. The colour code indicates the magnitude of $P_t$ and of the
  ratios, respectively. The couplings $g_A$ and $g_V$ have been chosen as 
  described in the text.} 
\end{center}
\end{figure}

Assuming a particle with mass of 120 GeV has been produced at an
$e^+e^-$ collider in association with a top-quark pair, it is
investigated if the cross section and its energy dependence as well as 
$P_t$ allow for the distinction of a SM Higgs $H^0$ (and hence a spin 0 
particle) from a spin 1 particle.
We therefore consider $e^+e^- \to t\bar{t} Z'$ production of a $Z'$ with mass 
$M_{Z'} = M_{H^0}=120$ GeV. The cross section is assumed to have the same
magnitude as expected for the production of a SM Higgs with this mass
value and a c.m. energy of 800 GeV, within an error of
10\%. This is the error on SM Higgs production 
to be expected for these parameter values at
an ILC. This condition leads to a range ${\cal P}$ of values 
$g_V$ and $g_A$, {\it i.e.}
\beq
g_V, g_A \in {\cal P} \quad \Longleftrightarrow \quad
\sigma_{t\bar{t}Z'} (g_V,g_A) = \sigma_{t\bar{t} H^0}  \pm 10\% \;.
\eeq
For these values, the top polarisation asymmetry is calculated and compared to
$P_t (t\bar{t}H^0) = 0.11$.  Fig.~\ref{fig:ptspin1} (upper left) shows
the values of $P_t (t\bar{t}Z')$ for the range of values ${\cal
  P}$. The white strips in the coloured area appear where $P_t
(t\bar{t}Z')$ differs from $P_t(t\bar{t} H^0)$ by less than $5
\sigma$. These regions are small. We conclude that for a particle produced in
association with a top quark pair with the same rate within the
experimental error as expected from $t\bar{t}H^0$ production,
the top polarisation asymmetry can be exploited to distinguish a spin 1
state from a spin 0 state if neither $g_V$ nor $g_A$ is close to 0. \s

Moreover, the ratio of cross sections at different c.m. energies can be
exploited. This is shown in Figs.~\ref{fig:ptspin1} upper right,
lower left and lower right. The ratio of the $t\bar{t} Z'$ cross
section at two different collider energies for three c.m. energy
combinations, respectively, is plotted for the parameter range ${\cal
  P}$ as defined above. All three ratio combinations are larger than 
the corresponding SM values and differ from them by more than
5$\sigma$. The SM values are  
\beq
\frac{\sigma_{H^0 t\bar{t}}(1000 \mbox{ GeV})}{\sigma_{H^0 t\bar{t}}(800
  \mbox{ GeV})} &=& 0.85 \qquad 
\frac{\sigma_{H^0 t\bar{t}}(1300 \mbox{ GeV})}{\sigma_{H^0 t\bar{t}}(800
  \mbox{ GeV})} = 0.61 \nonumber\\
\frac{\sigma_{H^0 t\bar{t}}(1300 \mbox{ GeV})}{\sigma_{H^0 t\bar{t}}(1000
  \mbox{ GeV})} &=& 0.72 \;.
\eeq 
The SM ratios are very different from the $t\bar{t}Z'$ production
ratios since the maximum of the SM cross section occurs at $\sim 750$ GeV
and hence all considered ratios in case of the SM Higgs are less than 1. For
$t\bar{t}Z'$ production the maximum lies at 1000 GeV or higher,
depending on the values of $g_V$, $g_A$.  Although these results of
course depend on the mass of the particle radiated from the top
quark pair and on the considered energies, they show that the top
polarisation asymmetry and/or ratios of cross sections are sensitive
to the spin value of the particle produced in association with $t\bar{t}$. \s

\begin{figure}[p]
\vspace*{-1cm}
\begin{center}
\includegraphics[width=10cm]{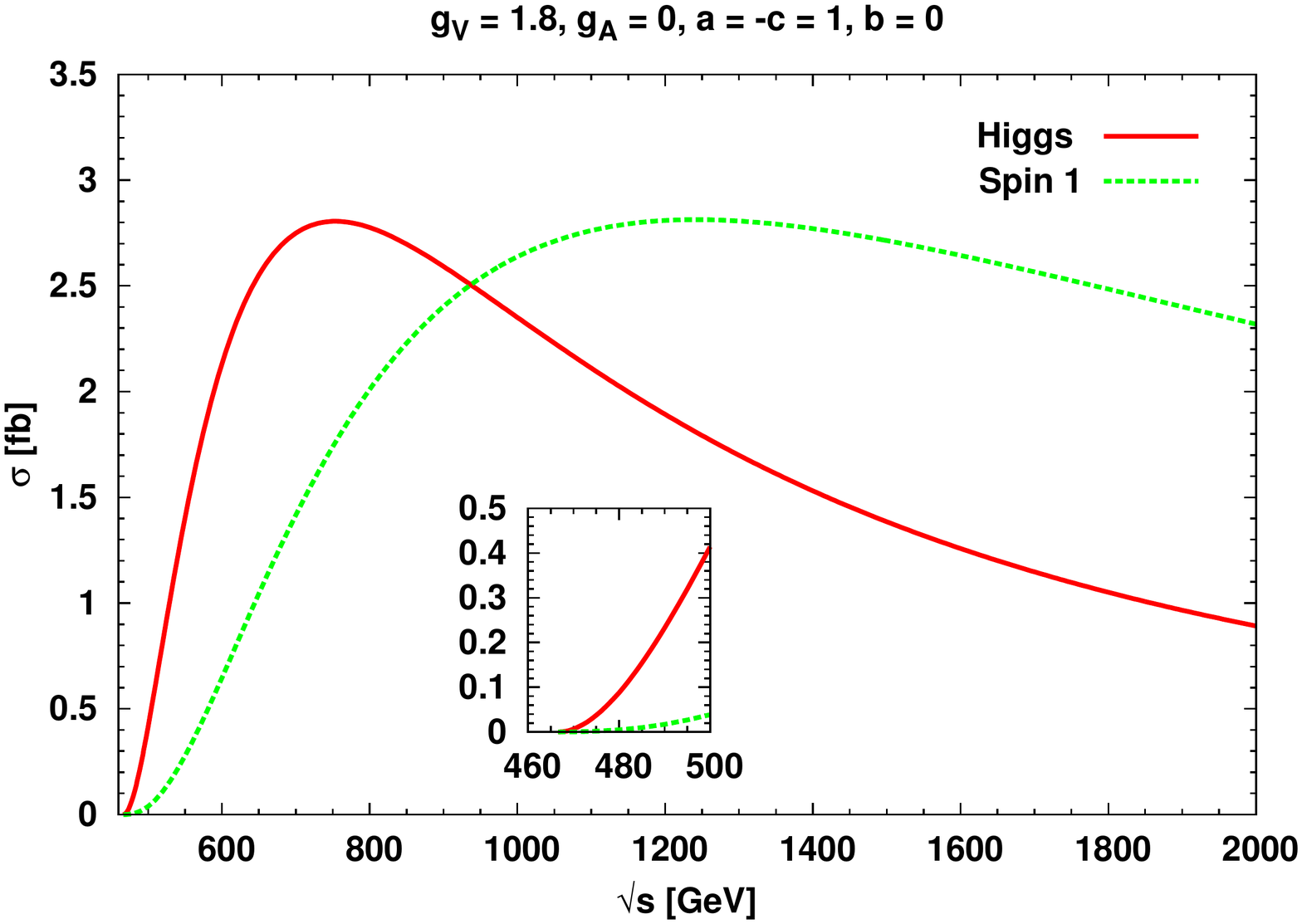}\\[-0.3cm]
\includegraphics[width=10cm]{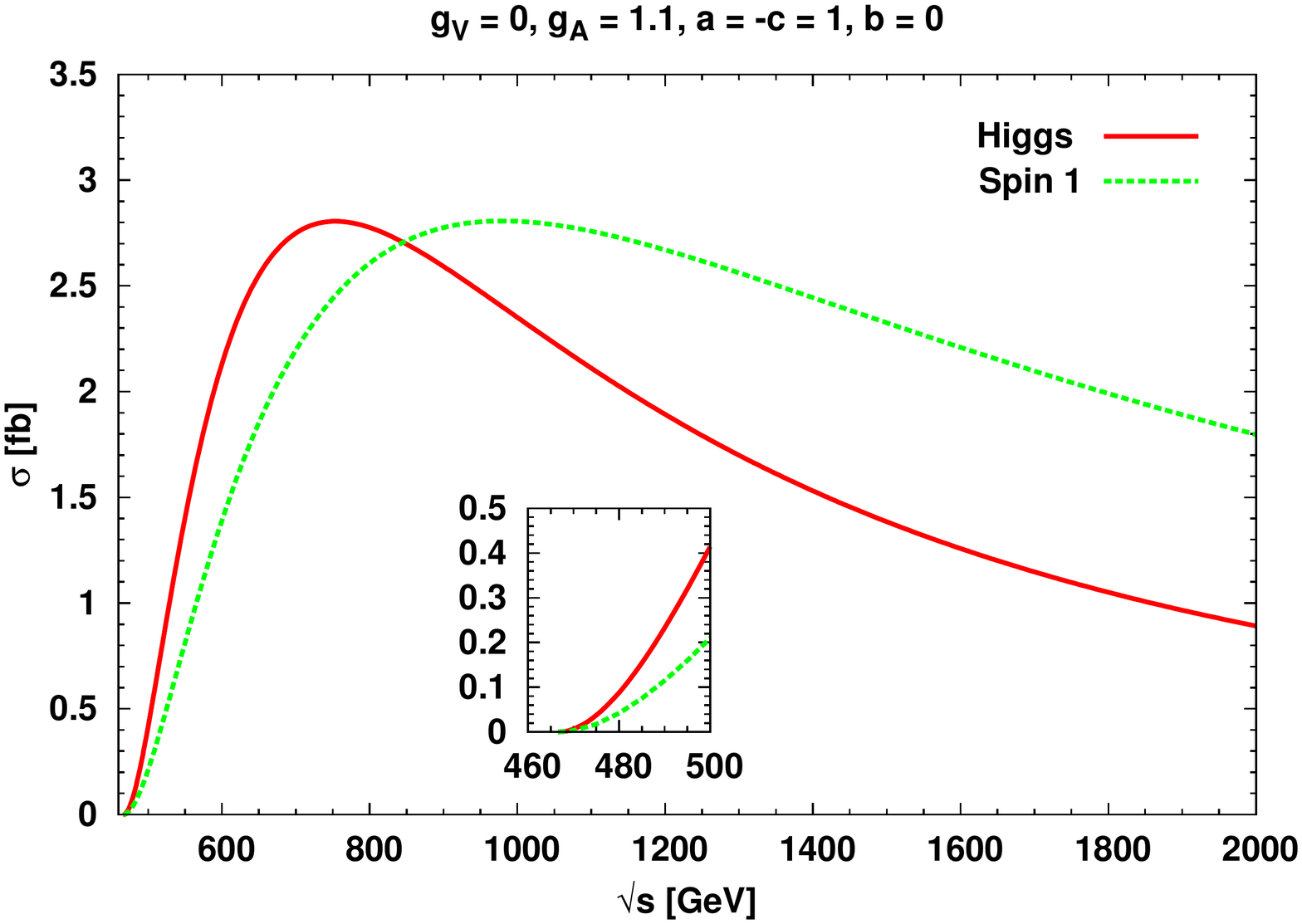}\\[-0.3cm]
\includegraphics[width=10cm]{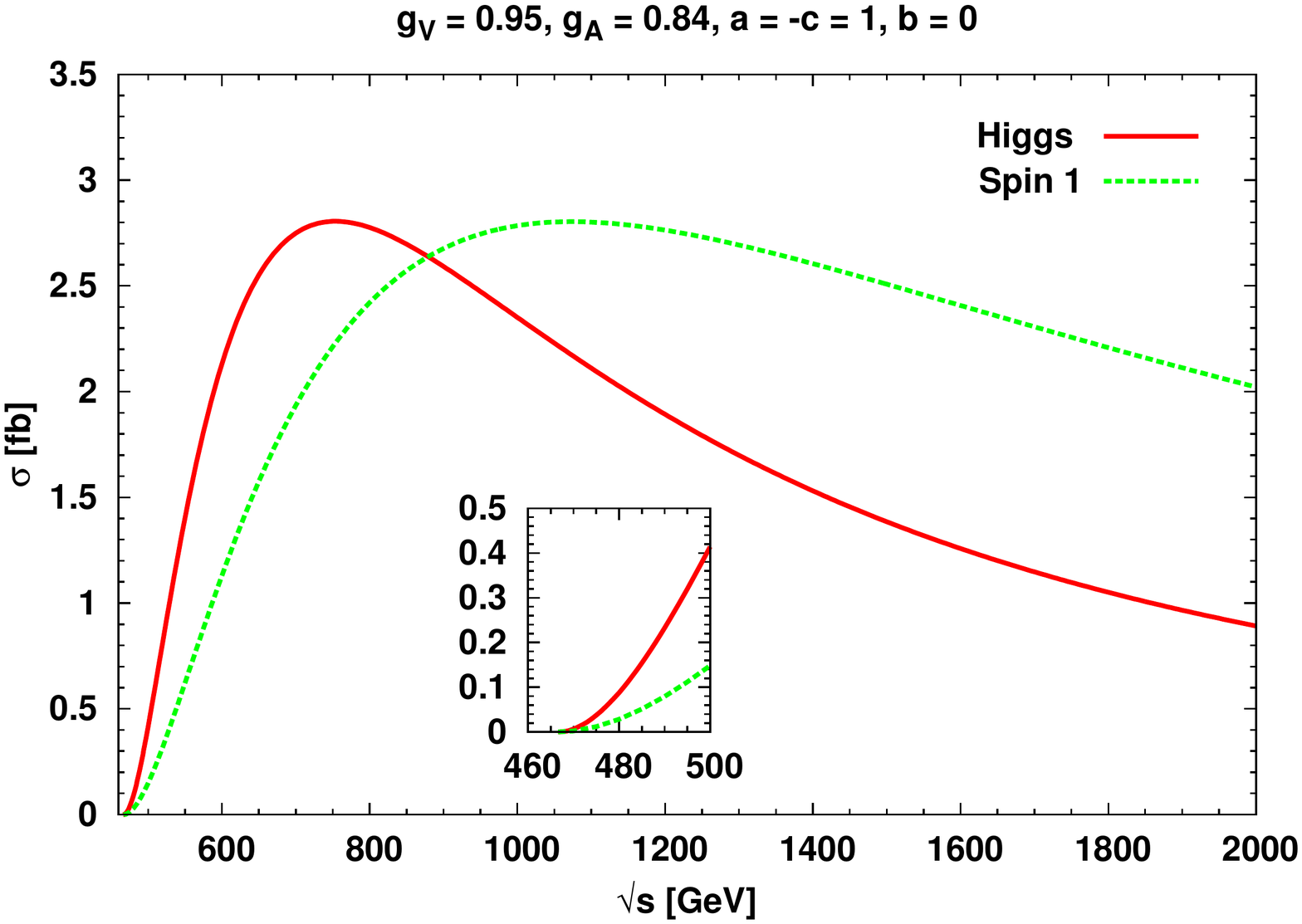}\\
\caption{\label{fig:spin1cpeven} Total cross section
  $\sigma(t\bar{t}H)$ (red/full) compared to $\sigma (t\bar{t}Z')$
  (green/dashed) as function of the 
  c.m. energy for $M_{Z'}=M_H=120$ GeV. The parameters $g_V \ne 0$
  (upper), $g_A \ne 0$ (middle) and $g_V,g_A \ne 0$ (lower)
  have been chosen such that the maximum values of the cross sections
  are the same.} 
\end{center}
\end{figure}

\begin{figure}[p]
\vspace*{-1cm}
\begin{center}
\includegraphics[width=10cm]{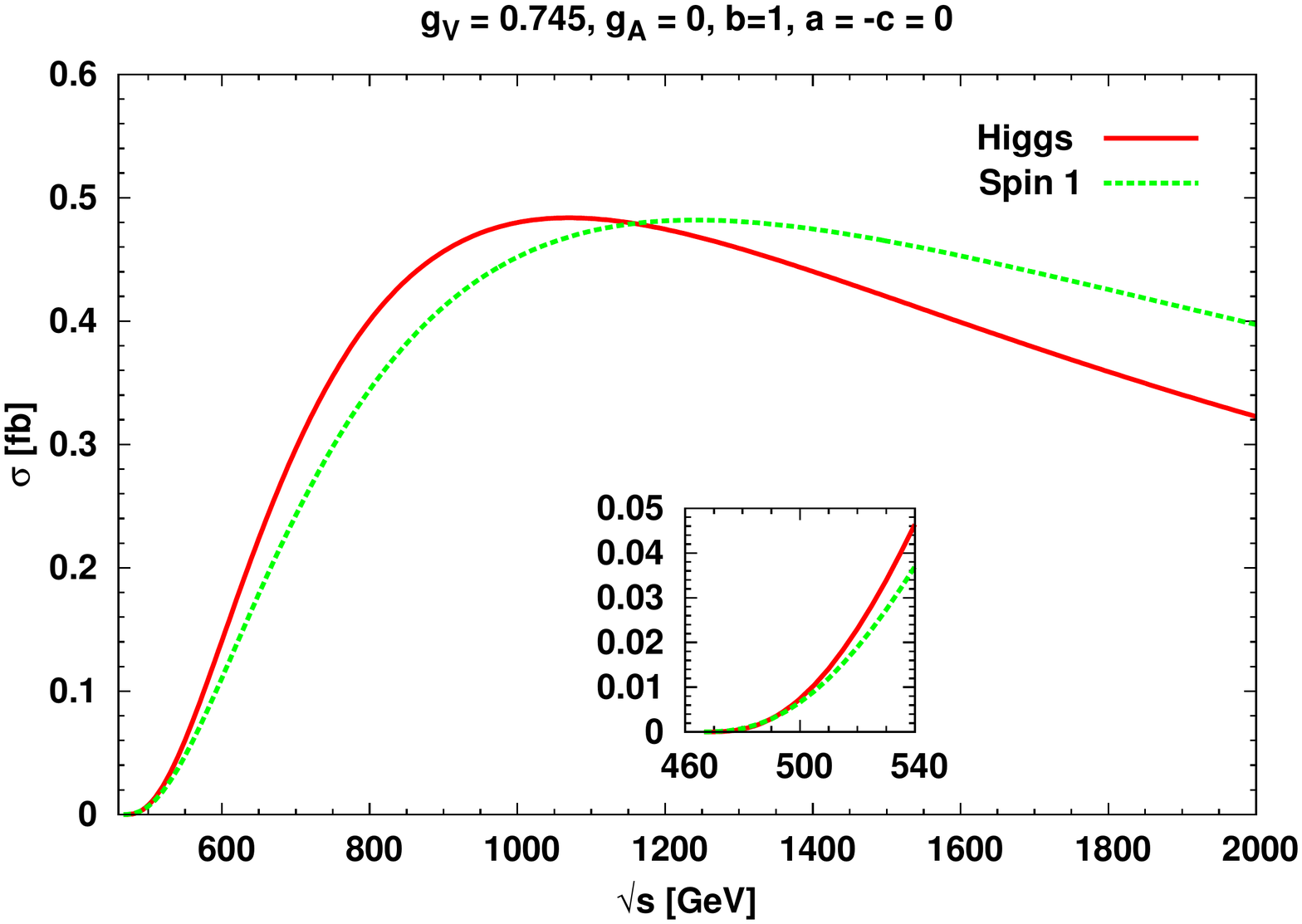}\\[-0.3cm]
\includegraphics[width=10cm]{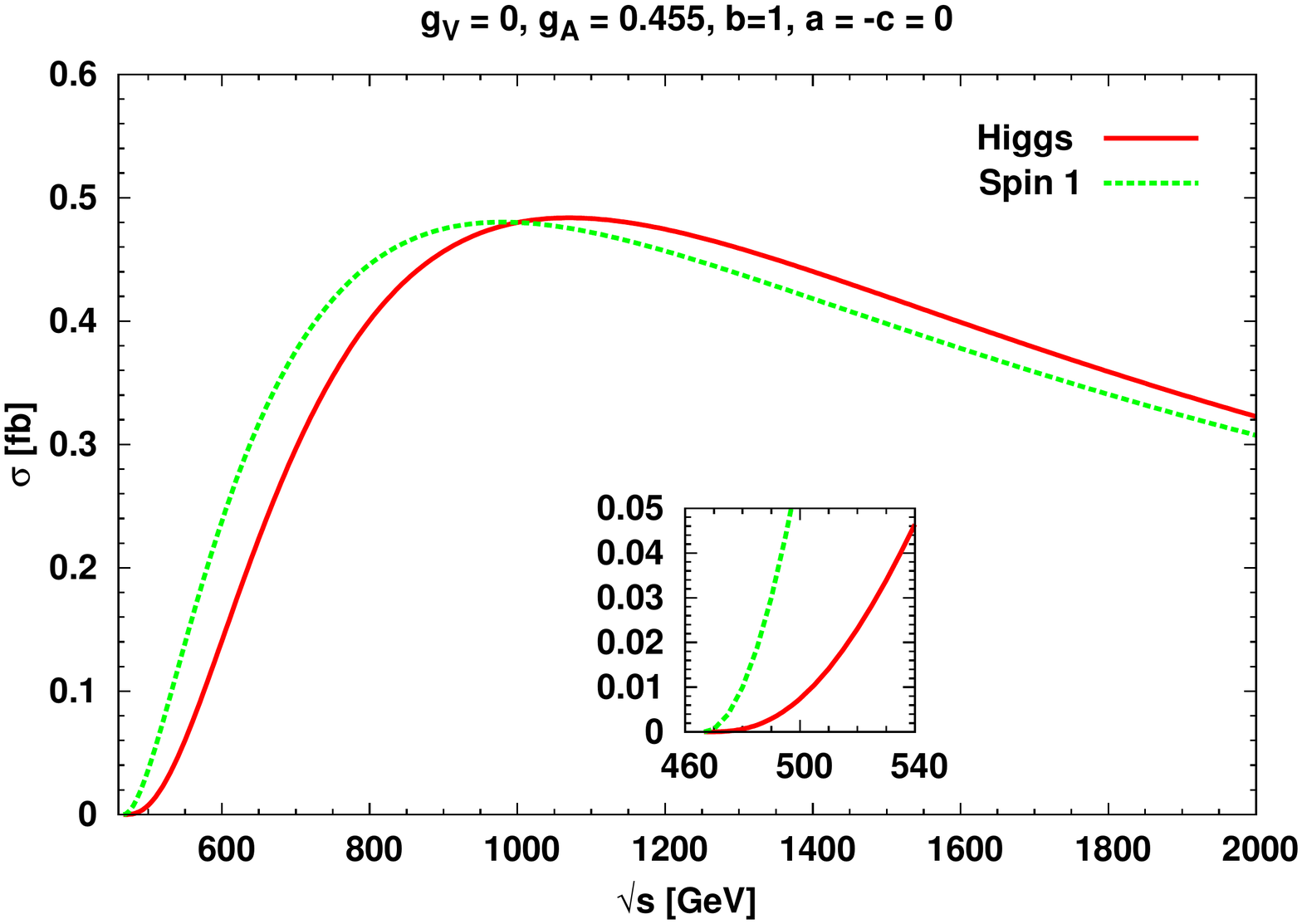}\\[-0.3cm]
\includegraphics[width=10cm]{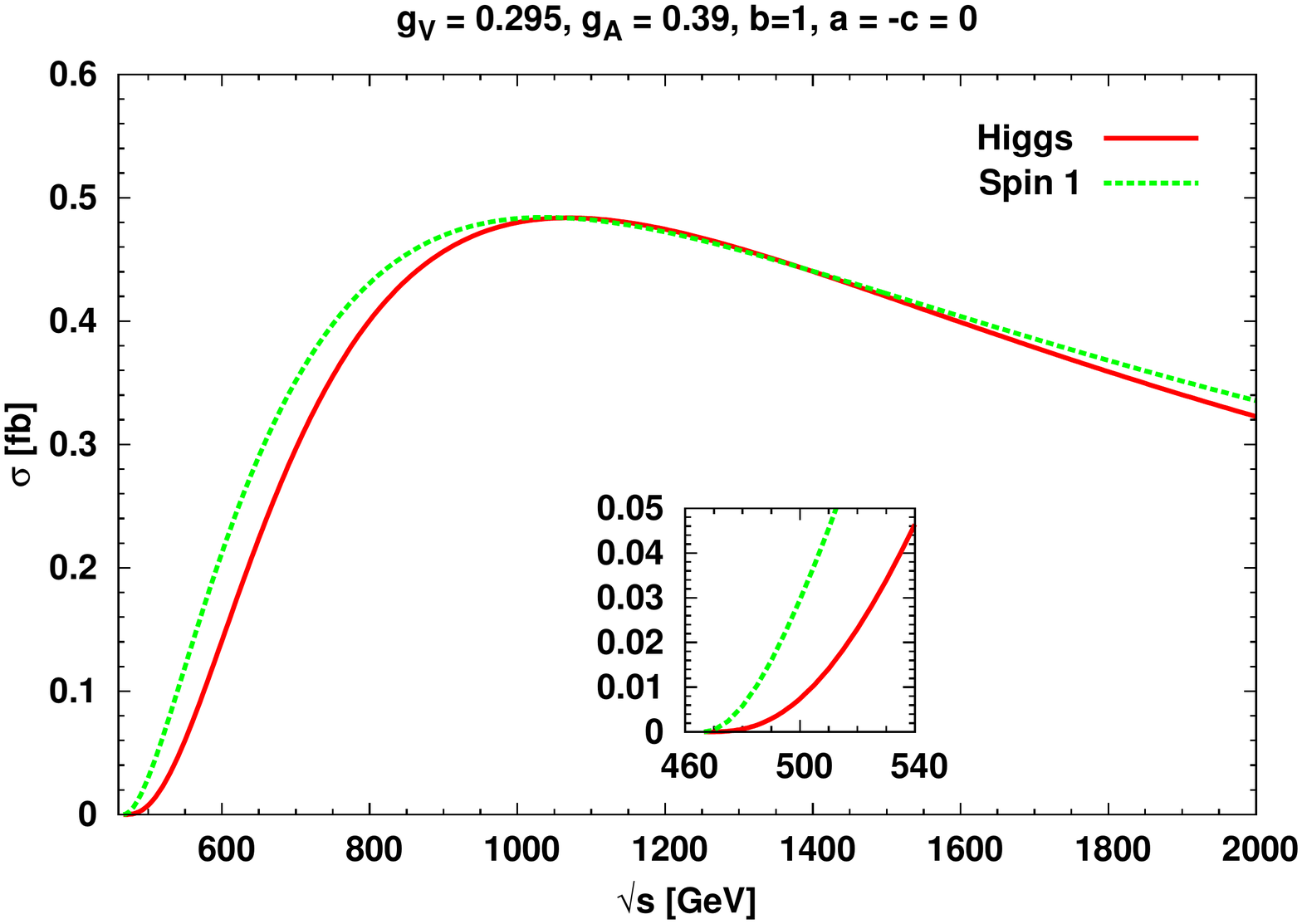}\\
\caption{\label{fig:spin1cpodd} Total cross section
  $\sigma (t\bar{t}A)$ (red/full) compared to $\sigma (t\bar{t}Z')$
  (green/dashed)  as function of the
  c.m. energy for $M_{Z'}=M_A=120$ GeV. The parameters $g_V \ne 0$
  (upper), $g_A \ne 0$ (middle) and $g_V,g_A \ne 0$ (lower)
  have been chosen such that the maximum values of the cross sections
  are the same.} 
\end{center}
\end{figure}

This statement is confirmed by the plots presented in
Figs.~\ref{fig:spin1cpeven} and
\ref{fig:spin1cpodd}. Figure~\ref{fig:spin1cpeven} shows the total  
cross section $\sigma(t\bar{t} Z')$ compared to the cross section of
scalar Higgs production in association with a top quark pair, $\sigma
(t\bar{t}H)$. The $H$ and $Z'$ masses have been chosen equal, $M_{Z'}
= M_H = 120$ GeV. For the Higgs coupling to the top quark pair the
maximum coupling strength $a=1$ has been adopted. The couplings $g_V$
and $g_A$ have been chosen within a parameter range ${\cal P'}$ such
that the maxima of the two cross sections in comparison are the same,
\beq
g_V, g_A \in {\cal P'} \quad \Longleftrightarrow \quad
\sigma_{t\bar{t}Z'}^{\mbox{\scriptsize max}} (g_V,g_A)  =
\sigma_{t\bar{t}H}^{\mbox{\scriptsize max}} \;. 
\eeq
Three cases are studied, where either $g_V$ or $g_A$ are non-zero and
where both $g_V$ and $g_A$ are non-zero. \s 

Figure~\ref{fig:spin1cpodd}
shows the comparison of $\sigma(t\bar{t} Z')$ with the associated
production of a pseudoscalar $A$ 

\noindent
with a top quark pair, $\sigma (t\bar{t} A)$, for $M_{Z'}=M_A=120$
GeV. For the CP-odd coupling parameter to the top quarks $b=1$ has
been assumed. The coupling parameters $g_V$, $g_A$ have been 
chosen as before within a parameter range ${\cal P''}$ such that the
maximum values of the cross sections are the same, 
\beq
g_V, g_A \in {\cal P''} \quad \Longleftrightarrow \quad
\sigma_{t\bar{t} Z'}^{\mbox{\scriptsize max}} (g_V,g_A)  =
\sigma_{t\bar{t}A}^{\mbox{\scriptsize max}} \;,
\eeq
and once again three types of coupling combinations $g_V$ and
$g_A$ are investigated. As can be inferred from the figures, the
curves for spin 1 production and CP-odd Higgs production are very
similar both in the threshold rise and in the continuum. This is to be
contrasted with the comparison of the curves for spin 1 production and
CP-even Higgs production. The threshold rise is not so much
different. But in the continuum they clearly differ. This reproduces
the difference in the ratios of cross sections observed above. \s


\section{\label{sec:concl} Summary} 
Once a Higgs boson has been discovered at the LHC, its CP properties
have to be determined. Should the Higgs boson not be CP-even but CP-odd or
a CP-mixed state, then the top sector represents an
ideal working ground to test its CP properties, since the various possible
CP-parts of the Higgs boson couple democratically to the top quarks.
This is in contrast to the Higgs coupling to gauge bosons which
projects out the CP-even part of the coupling and has only a small
admixture of CP-odd parts from loop diagrams. In this paper,
we investigated the production of a spin 0 state with
arbitrary model-independent CP properties in association with a top
quark pair at a future $e^+e^-$ linear collider. The CP properties of
the Higgs coupling to the top quarks have been parametrized in a
model-independent way by a parameter $a$ for a CP-even Higgs, by a
parameter $b$ for a CP-odd Higgs and by simultaneously non-vanishing
$a$ and $b$ for a CP-mixed state. These parameters can be determined
by a measurement of the total cross section, the polarisation
asymmetry of the top quark and the up-down asymmetry of the antitop
quark with respect to the top-electron plane. The former two
observables are CP-even and can be exploited to distinguish a CP-even
from a CP-odd Higgs boson.  Since the up-down asymmetry $A_\phi$ is
CP-odd, it can be exploited directly and unambiguously to test CP violation. \s

The sensitivities to $a$ and $b$ have been studied in each observable
separately before investigating the combination of all three observables. 
We found that the total cross section is most sensitive to $a$ and to
some extent to $b$. The observables $P_t$ and $A_\phi$ do not exhibit
much sensitivity to $a$ and $b$, although polarisation of the initial
$e^\pm$ beams slightly improves the sensitivity in case of $P_t$.  
The combination of all three observables, however, remarkably reduces
the error on $a$ for polarised $e^\pm$ beams. At 800 GeV c.m. energy
this is due to the mutual interplay of $\sigma$ and $P_t$ in
constraining the parameter range of $a$ and $b$. At multi-TeV energies
$A_\phi$ is larger, and it is now the interplay of all three
observables which further reduces the error. \s

We also investigated to what extent the radiation of an arbitrary
spin 1 particle from the top quark pair, which like the Higgs boson
does not couple to $e^\pm$, can be distinguished from a spin 0
state with the same mass. It turns out that the top polarisation
asymmetry and ratios of cross sections at different collider energies
represent good observables to tell a spin 0 from a spin 1 state. \s

In summary, associated Higgs production with a $t\bar{t}$ pair at a
future $e^+e^-$ collider allows for the determination of the spin and
CP properties of the Higgs boson in a model-independent way.
Combining the total cross section and its energy dependence, the
top polarisation asymmetry and the up-down asymmetry considerably
helps to reduce the errors on the coupling parameters of the Higgs
Yukawa coupling which reveal the CP nature of the Higgs boson.

\section*{Appendix}
\subsection*{Cross sections for initial beam polarisation}
The cross sections $\sigma_0$ and $\sigma_1$ as defined in
Eq.~(\ref{eq:defsig0sig1}) get modified for initial beam
polarisation. The polarisation acts differently on the contributions
to the cross sections from the various diagrams. The changes can be 
summarized by replacing the coupling factors in the cross
sections. Denoting by $P_{e^-}$ and $P_{e^+}$ the degree of longitudinal
polarisation 
of  the electron and positron, respectively, the changes to be applied for
the couplings appearing in $\sigma = 2 \sigma_0$
Eq.~(\ref{ttHxsection}) and $\sigma_1$ Eq.~(\ref{eq:sigma1}) are
\beq
Q_e^2Q_t^2 &\rightarrow& Q_e^2Q_t^2 \left(1-P_{e^+} P_{e^-} \right)
\nonumber \\
2Q_e Q_t v_e &\rightarrow&
2Q_eQ_t  \left[v_e \left(1-P_{e^+} P_{e^-}\right) + a_e
  \left(P_{e^+} - P_{e^-} \right)\right] 
\\
(v_e^2+a_e^2) &\rightarrow &
\left[(v_e^2 + a_e^2) \left(1-P_{e^+} P_{e^-} \right) + 2 v_e a_e
  \left(P_{e^+} - P_{e^-}\right)\right] \nonumber \;.
\eeq
The up-down asymmetry originates from the interference term
between the diagrams where the Higgs is radiated from the top line and
the one where the Higgs is radiated from the $Z$ boson. The
denominator of $A_\phi$ is simply given by the total cross section
{\it cf.} Eq.~(\ref{eq:aphi}). The differential cross section for
the interference term $\sim bc$ reads in terms of  the scaled Mandelstam
variables $S_1,S_2,T_1,T_2$\\ 
\beq
\frac{d \sigma_{bc}}{dS_1 dS_2 dT_1 dT_2} &=& \frac{3 \alpha^2}{8\pi^2 s^3}
 \frac{bc \; g_{ttH} g_{ZZH} \sqrt{h_t/s}}{h_Z (S_1 -h_Z)
  (S_2-h_t)(-1+S_1+S_2-h_t-h_\Phi) } \times
\nonumber \\
&& \Big\{ \frac{Q_e Q_t a_t a_e}{(h_Z -1)} D_1^\Phi
+ \frac{2a_e v_e a_t v_t}{(h_Z-1)^2} D_2^\Phi
+ \frac{a_t^2(a_e^2+v_e^2)}{(h_Z-1)^2} D_3^\Phi \Big \} \,
\epsilon^{p_1 p_2 q_a q_b} \;, 
\label{eq:updownas}
\eeq
with $h_X = m_X^2/s$ and the form factors given by
\beq
D_1^\Phi &=& (h_\Phi+1-S_1)(h_\Phi-2h_Z -1 +S_1) \\
D_2^\Phi &=& (1-S_1)^2-h_\Phi^2 \\
D_3^\Phi &=& (-h_\Phi-2h_t-1+S_1+2S_2) (h_\Phi-1+S_1-2T_2) \;.
\eeq
The scaled Mandelstam variables in terms of the reduced energies
$x_1,x_2$, the azimuthal angle $\chi$ of the antitop, the polar angle
$\theta$ of the top quark, the angle $\theta_{12}$ between top and
antitop and the  velocities $\beta_{1,2}=\sqrt{1-4h_t/x_{1,2}^2}$ read
\beq
S_1 &=& 1 - h_\Phi - x_3 \\
S_2 &=& 1+h_t-x_1 \\
T_1 &=& h_t - \frac{1}{2} x_1 (1-\beta_1 \cos \theta) \\
T_2 &=& h_\Phi - \frac{1}{2} x_1
[  x_3-\sin\theta\sin\theta_{12}\cos\chi x_2\beta_2-
\cos\theta (x_1\beta_1+x_2\beta_2\cos\theta_{12}) ] \;.
\eeq
In case of polarised $e^\pm$ beams, the following replacements for
the coupling factors have to be made
\beq
Q_e Q_t a_e &\to& Q_e Q_t \left[a_e (1-P_{e^+} P_{e^-}) + v_e
  (P_{e^+} - P_{e^-} ) \right] \nonumber \\
2 a_e v_e &\to& [2 v_e a_e (1-P_{e^+} P_{e^-} ) +
   (a_e^2+v_e^2)  (P_{e^+} - P_{e^-} )] \\
(a_e^2+v_e^2) &\to& [ (a_e^2+v_e^2)  (1-P_{e^+} P_{e^-}) + 
2 a_e v_e (P_{e^+} - P_{e^-} ) ] \;. \nonumber
\eeq

\section*{Acknowledgments}

\noindent
RG and SDR wish to acknowledge support from the Department of
Science and Technology, India under the J.C. Bose Fellowship scheme
under grant nos. SR/S2/JCB-64/2007 and SR/S2/JCB-42/2009,
respectively. CH and MM acknowledge support from the Deutsche
Forschungsgemeinschaft via the Sonderforschungsbereich/Transregio
SFB/TR-9 Computational Particle Physics. 


\end{document}